\documentclass[%
reprint,
 amsmath,amssymb,
aps,
pra,twocolumn,Sqrt
]{revtex4}
\usepackage{graphicx}
\usepackage{dcolumn}
\usepackage{bm}
\usepackage[hidelinks]{hyperref}
\hypersetup{colorlinks = true,
  urlcolor = blue,
  linkcolor = blue,
  citecolor = blue
}
\usepackage{bigints}

\usepackage{url}

\usepackage[normalem]{ulem}

\newcommand{\cbeims}[1]{{\color{red}#1}}

\begin{document}


\title{Position as an independent variable and the emergence of\\  the $1/2$-time fractional derivative in quantum mechanics}

\author{Marcus W. Beims}
\email{mbeims@fisica.ufpr.br}
\author{Arlans J. S. de Lara}
 \email{arlansslara@gmail.com}
\affiliation{Departamento de Física, Universidade Federal do Paraná, 81531-990, Curitiba, PR, Brazil
}%

\date{\today}
\begin{abstract} 
Using the position as an independent variable, and time as the dependent variable, we derive the function ${\cal P}^{(\pm)}=\pm\sqrt{2m({\cal H}-{\cal V}(q))}$, which generates the space evolution under the potential ${\cal V}(q)$ and Hamiltonian ${\cal H}$. No parametrization is used. Canonically conjugated variables are the time and minus the Hamiltonian  ($-{\cal H}$).   While the classical dynamics do not change,  the corresponding quantum operator  ${\cal \hat P}^{(\pm)}$  naturally leads to a $1/2-$fractional time evolution, consistent with a recent proposed spacetime symmetric formalism of the quantum mechanics. Using Dirac's procedure, separation of variables is possible, and while the two-coupled position-independent Dirac equations depend on the $1/2$-fractional derivative, the two-coupled time-independent Dirac equations (TIDE) lead to positive and negative shifts in the potential, proportional to the force.   Both equations couple the ($\pm$) solutions of ${\cal \hat P}^{(\pm)}$ and the kinetic energy ${\cal K}_0$ (separation constant) is the coupling strength. Thus, we obtain a pair of coupled states for systems with finite forces, not necessarily stationary states. The potential shifts for the harmonic oscillator (HO) are $\pm\hbar\omega/2$, and the corresponding pair of states are coupled for ${\cal K}_0\ne 0$. No time evolution is present for ${\cal K}_0=0$, and the ground state with energy $\hbar\omega/2$ is stable. For  ${\cal K}_0>0$, the ground state becomes coupled to the state with energy $-\hbar\omega/2$, and \textit{this coupling} allows to describe higher excited states in the HO. Energy quantization of the HO leads to the quantization of  ${\cal K}_0=k\hbar\omega$ ($k=1,2,\ldots$).  For the one-dimensional Hydrogen atom,  the potential shifts become imaginary and position-dependent. Decoupled case ${\cal K}_0=0$ leads to plane-waves-like solutions at the threshold. Above the threshold (${\cal K}_0>0$), we obtain a plane-wave-like solution, and for the bounded states (${\cal K}_0<0$), the wave-function becomes similar to the exact solutions but squeezed closer to the nucleus.
\end{abstract}


\maketitle

\section{Introduction}
\label{intro}

The accurate role and identification of dependent and independent physical quantities, and their connection, are of crucial relevance in describing physical systems. In Newton's equation of motion, while the position of a particle is the dependent variable, the time has been treated as an independent parameter. However, for many centuries researchers tried to understand the benefit of treating time as a dynamic variable in the classical equations of motion. Such treatments lead naturally to considering energy and time as conjugated variables. In this context, earlier works \cite{lanczos49,synge60,haar61,hja62,marius00} consider the time $t$ (and sometimes the energy) as an \textit{additional} dependent variable, meaning that together with the velocity and position coordinates ($q,\dot q = dq/dt$), we have the time coordinate $t(\tau)$, where $\tau$ is inserted as parameter. It is a parametrization of the equations of motion and is of significant interest in relativistic mechanics. \\

Different from the above approaches, we propose that \textit{space} is the \textit{independent} variable, and \textit{time}  the \textit{dependent} parameter, with \textit{no extra dimensions or parametrization}. Newton's law is transformed as
\begin{equation}
F=m\ddot q=-m\frac{t''}{t'^3},
\label{mddot}
\end{equation}
$m$ is the particle's mass, and the prime means derivative relative to $q$ ($t'=dt/dq$). We start obtaining the Lagrangian, and using Hamilton formalism we derive the "Momentumian" (units of momentum), which corresponds to the Hamiltonian in the usual classical description. The time ($t$) and minus the total energy ($-{\cal H}$) become the conjugated variables. The canonical equations of motion are obtained from the Momentumian, responsible for generating the canonical \textit{space} translation, in analogy to the Hamiltonian, which generates the \textit{time} translation in the usual mechanical interpretation. For time-independent potentials ${\cal V}(q)$, we derive the \textit{Momentumian} 
\begin{equation}
{\cal P}^{(\pm)}(t,{\cal H};q) = \pm\sqrt{2m({\cal H}-{\cal V}(q))},
\label{P}
\end{equation}
where we recognize that ${\cal H}-{\cal V}(q)= {\cal K} = m/(2t'^2)$ is the kinetic energy written as a function of the velocity of the time $t'$. Calligraphic fonts are used when the physical quantities are calculated, considering position as the independent parameter.    
The canonical quantization of ${\cal P}(t,{\cal H};q)$  leads precisely to the quantum operator in the spacetime-symmetric (STS) formalism proposed by Dias and Parisio \cite{diasparisio}, together with its more recent discussion about the interpretation of the meaning behind the classical parameter $q$ in the quantum evolution~\cite{araujo2023}.

The square root of quantum operators produces $1/2$-fractional time derivatives. Therefore, the description of the position as an independent variable \textit{naturally}  leads to the appearance of the $1/2$-fractional time derivative, which appears in the quantum STS formalism. No \textit{ab initio} fractional assumption is demanded, as usual, in the context of fractional applications \cite{laskin1,laskin2,laskin3,laskin,tarasov2013,tarasov}. 

Separation of variables becomes possible using Dirac's procedure, leading to two-coupled $1/2$-order position-independent Dirac equations (PIDE) and two-coupled first-order time-independent Dirac equations (TIDE). The separation constants $P^+$ and $P^-$ couple the two equations and are quantified by the kinetic energy ${\cal K}_0=P^+P^-/2m$. While the PIDE is trivially solved regarding Mittag-Leffler functions, the TIDE depend on the specific physical system. Rewriting the two first-order TIDE as one second-order equation, we obtain a time-independent Schrödinger equation (TISE) with an effective potential 
\begin{equation}
\label{VeffI}
{\cal V}^{(\pm)}_{\mbox{\tiny eff}}= {\cal V}(q)\mp \frac{\hbar}{2\sqrt{2m{\cal V}(q)}}{\cal F}(q),
\end{equation}
where ${\cal F}(q)=-d{\cal V}(q)/dq$ is the force.  For the uncoupled case ${\cal K}_0=0$, there is no time evolution, and the corresponding time-independent Schrödinger equation is ${\cal H}^{(\pm)}_{\mbox{\tiny eff}}\chi(q)=0$, having zero energy. The prohibition of having states with zero energies in bounded quantum systems is compensated inside ${\cal H}^{(\pm)}_{\mbox{\tiny eff}}$ with the positive and negative shifts in the potential (\ref{VeffI}). Excited states of the HO are obtained using ${\cal  K}_0=k\hbar\omega$ (with $k=1,2,\ldots$). For the one-dimensional Hydrogen atom, it leads to plane-waves-like solutions at the threshold. For the coupled Dirac equations, forward (${\cal K}_0>0$) and backward  (${\cal K}_0<0$) in time evolution generates continuum and bounded states, respectively. They differ from the exact usual by being squeezed closer to the nucleus.

In the quantum world, our interpretation, as the STS formalism \cite{diasparisio}, brings us to an old and still not completely solved problem: the role and interpretation of time in quantum mechanics (QM). Though earlier works \cite{pauli33,bohm61,razavy67,razavy69,allcock69-1,allcock69-2,allcock69-3,kijo74} already considered the possibility to have a time operator in QM, increasing related studies and debates in recent years \cite{briggs1,briggs2,briggs3,hoge10} became fashionable, mainly in the context of the time of arrival in QM \cite{tate96,muga97,muga98,muga99,muga00,andersontime,galapon04,galapon05-1,galapon05-2,secao122,galapon06,galapon08,galapon09,galapon18,galapon22,arlans23-1}, which can be checked experimentally (see \cite{secao122} and references therein). In this context, the main issue is that distributions of the arrival time should be derived and interpreted. In the usual QM approach, $\psi(q;t)=\langle q|\psi(t)\rangle$ gives the probability amplitude of finding the particle within $q$ and $q + dq$, given that the time of detection is {\it exactly} $t$. Alternatively, we could also ask about  $\phi(t;q)=\langle t|\psi(q)\rangle$, which gives the probability amplitude of finding the particle at times $t$ and $t + dt$, given that the position of detection is {\it exactly} $q$. For more details, see \cite{diasparisio}. \\

In the present paper, we limit the discussion to the one-dimensional case with time-independent potential. While in Sec.~\ref{momen} we derive the classical Momentumian and discuss some textbook examples, Sec.~\ref{Qversion} analyses the quantum case. We show examples of the free particle, harmonic oscillator and one-dimensional Hydrogen atom. Section \ref{conclusions} summarises our results and furnishes future developments.

\section{Conjugated variables and the Momentumian}
\label{momen}

We assume that instead of measuring the position as a function of time $q(t)$, we measure time as a function of the position $t(q)$; in this approach, the position is the independent variable. For consistency, the final physical solution must be the same. Using the relation between second derivatives of  inverse functions, namely $\ddot q=-\dot q^3t''=-t''/t'^3$, Newton's second law, Eq.~(\ref{mddot}), is written as
\begin{eqnarray}
-\frac{m}{2}\frac{d}{dq}\left[\frac{\partial (1/t')}{\partial t'} \right] &=& {\cal F}(q).
\label{t22}
\end{eqnarray}
 We notice that  $\ddot q=-t''/t'^3$ is valid if $t$ is twice differentiable and $t'$  is nonzero. The latter means we should act carefully when infinite velocities $\dot q$ become relevant. The general solution for Eq.~(\ref{t22}) is 
\begin{equation}
    t(q) = t_0 \pm \bigints_{q_0}^q\sqrt{\frac{m}{2\left(\frac{m}{2 t_0^{'2}} +  \int_{q_0}^{\bar{q}} {\cal F}(\bar{\bar{q}})\mathrm{d} \bar{\bar{q}}\right)}}d \bar{q}.
\label{tqfall}
\end{equation}

To obtain the canonical equations of motion, we start with the functional
\begin{equation}
{\cal S}=\int_{q_i}^{q_f} {\cal L}(t,t';q)\,dq,
\label{EQF}
\end{equation}
where $t' = dt/dq$, whose corresponding Euler-Lagrange equation is 
\begin{equation}
\frac{\partial {\cal L}}{\partial t}-\frac{d}{dq}\left(\frac{\partial {\cal L}}{\partial t'}\right)=0,
\label{EQLF}
\end{equation} 
where ${\cal L}(t,t';q)$ must be found and interpreted. Equation (\ref{EQLF}) provides the condition which $t(q)$ has to satisfy so that ${\cal S}$ is minimal along the path $q_i\to q_f$. The solution of  Eq.~(\ref{EQLF}) should be consistent with the usual Lagrangean, therefore 
\begin{eqnarray}
S&=&\int_{t_i}^{t_f} L(q,\dot{q};t) dt= \int_{t_i}^{t_f} \left[\frac{m}{2}\dot q^2 - V(q)\right]\,dt\cr
& & \cr
                      &=& \int_{t_i}^{t_f}  \left[\frac{m}{2}\dot q^2 - V(q)\right]\frac{dt}{dq}\,dq= \int_{q_i}^{q_f}  \left[\frac{m}{2t'} - V(q)\,t'\right]\,dq.\nonumber
\end{eqnarray}
The last term we recognize to be ${\cal S}$ from Eq.~(\ref{EQF}) if
\begin{equation}
{\cal L}(t,t';q) = \frac{m}{2t'} - {\cal V}(q)\,t',
\label{Sf2}
\end{equation}
and the associated Euler-Lagrange equation leads to
$- {m\,t''}/{t'^3} = {\cal F}$,
as expected. The essential issue is to realize that in this interpretation, the minimal principal which generates the dynamics is obtained from the Lagrangian (\ref{Sf2}), which has units of momentum.   \\

As usual, the  conjugated variables, say $(t,h)$, are obtained from
\begin{equation}
h = \frac{\partial {\cal L}(t,t';q)}{\partial t'} =-\frac{m}{2t'^2}-{\cal V}(q)=-{\cal H},
\label{cE}
\end{equation}
${\cal H}$ is the usual total energy, but written in terms of $t'$. This equips us with the Poisson brackets $\{t,h\} = -\{t,{\cal H}\} = 1$. Substituting Eq.~(\ref{cE}) in Eq.~(\ref{EQLF}) we get
\begin{equation}
h'=\frac{dh}{dq} = \frac{\partial{\cal L}}{\partial t} = -{\cal H}'.
\end{equation} 
The Momentumian ${\cal P}(t,{\cal H};q)$ is generated by the Legendre transformation
\begin{eqnarray}
{\cal P}(t,{\cal H};q)&=&t'h - {\cal L}(t,t';q),\quad \mbox{or}\cr
& & \cr
{\cal P}(t,{\cal H};q)&=&-t'{\cal H} - {\cal L}(t,t';q),
\label{Leg}
\end{eqnarray}
and we arrive at the \textit{canonical equations of motion} for $(t,-{\cal H)}$, namely
\begin{eqnarray}
t'=-\frac{\partial {\cal P}}{\partial {\cal H}},\quad {\cal H}'=\frac{\partial {\cal P}}{\partial {t}},\quad \frac{\partial {\cal L}}{\partial {q}} = - \frac{\partial {\cal P}}{\partial {q}}.
\end{eqnarray}
Using Eqs.~(\ref{Sf2}) and (\ref{Leg}) we obtain the \textit{Momentumian}  
\begin{eqnarray}
{\cal P}^{(\pm)}(t,{\cal H};q)&=&\pm\sqrt{2m({\cal H}-{\cal V}(q))} , 
\label{PP}
\end{eqnarray}
where ${\cal H} -{\cal V}(q)={\cal K}=m/t'^2$ is always positive for real $t'$. The canonical equations of motion become
\begin{eqnarray}
  \label{t}
  t'&=&\mp\sqrt{\frac{m}{2({\cal H}-{\cal V}(q))}} = \frac{m}{{\cal P}^{(\mp)}(t,{\cal H};q)}, \\
& & \cr
& & \cr
{\cal H}'&=& \pm\sqrt{\frac{m}{2({\cal H}-{\cal V}(q))}}\left(\frac{\partial {\cal V}(q)}{\partial t}\right) = 0.
\end{eqnarray}
Thus, for time-independent potentials, we always have $d{\cal H}/dq=0$, ${\cal H}={\cal H}_0$ is the cyclic variable and a \textit{constant of motion in space}, as can be easily checked. Another statement is that $t'$ always diverges at the turning points for which $[{\cal H}_0-{\cal V}(q)]=0$. Thus, time changes faster and faster as approaching the turning points.

Finally, the solution for $t(q)$ can be obtained by solving Eq.~(\ref{t})
  \begin{eqnarray}
    t(q) &=& t_0\mp \int_{q_0}^{q_f}\frac{m}{\sqrt{2m({\cal H}_0-{\cal V}(q))}}dq, \nonumber\\
    &=& t_0\mp \int_{q_0}^{q_f}\frac{m}{{\cal P}^{(\mp)}(t,{\cal H};q)}dq.
\end{eqnarray}
Thus, due to the signs $\mp$, we always have two solutions (branches) for the time, which lead to the \textit{same} physical solution. Depending on $t_0$, we may have negative times,  which can be avoided by choosing large enough values of $t_0$ (this is what we assume here for the considered times). As we will demonstrate afterwards, the relevant point is not if the time is negative or positive but if the velocity of the time $t'$ is positive or negative. Furthermore, we call to attention that the plus (minus) sign in the Momentumian from Eq.~(\ref{PP}) corresponds to negative (positive) velocities of the time, Eq.~(\ref{t}). In other words, consider the time at two positions, $t(q_1)=t_1$ and  $t(q_2)=t_2$, the velocity of the time is then $t'\sim\frac{t_2-t_1}{q_2-q_1}.$ Therefore: \\
$\bullet$ For ${\cal P}^+(t,{\cal H}_0;q)>0$ we have from Eq.~(\ref{t}) that $t'<0$. This is possible when  $q_2>q_1$ and $t_2<t_1$, or when $q_2<q_1$ and $t_2>t_1$. Thus, when space increases (decreases), time decreases (increases). In other words, motion to the right (left) with decreasing (increasing) times. In other words, space and time have inverted directions in their axis.\\
$\bullet$ For ${\cal P}^-(t,{\cal H}_0;q)<0$ we have from Eq.~(\ref{t}) that $t'>0$. 
This is possible when  $q_2>q_1$ and $t_2>t_1$, or when $q_2<q_1$ and $t_2<t_1$. Thus, motion to the right (left) with increasing (decreasing) times. In other words, space and time have the same directions in their axis.\\

For later purposes, we mention that in the classical forbidden region, $[{\cal H}_0-{\cal V}(q)]<0$, the quantities ${\cal P}^{\pm},t,t'$ and $t''$ become complex, even for a real initial time $t_0$. Since $t'$ is a pure complex number, we can write the time evolution as $t(q)=t_0+it_I(q)$,  so that the time difference between two positions will be
\begin{equation}
    t(q_f) - t(q_i) = [t_0 + i t_I(q_f)] - [t_0 + i t_I(q_i)] = i [ t_I(q_f) -  t_I(q_i)],
    \nonumber
  \end{equation}
which is a pure complex number.

Furthermore, from ${\cal K}=m/t'^2$ we observe that $t'(t;q)= t'(-t;-q), t'(t;q)= -t'(-t;q),$ and $t'(t;q)= -t'(t;-q)$. Thus, we can interchange between  ${\cal P}^+(t,{\cal H};q)$ and ${\cal P}^-(t,{\cal H};q)$ by changing $t\to-t$ \textit{or} $q\to -q$. 

\subsection{Examples}
\label{examples}

Now we discuss some simple textbook examples to understand the time behaviour better.

\noindent
$\bullet$  \textit{For the  free particle case} the canonical equations of motion are $t'=\mp\sqrt{m/(2{\cal H})}$ and ${\cal H}'=0$,  the quantity ${\cal H}={\cal H}_0$ is constant in space and $t(q)=t_0\mp\sqrt{m/(2{\cal H}_0)}(q-q_0)$. This solution agrees with a direct integration of Eq.~(\ref{t22}) with initial condition $t'_0=\mp\sqrt{m/(2{\cal H}_0)}$. Time increases (decreases) linearly as a function of $q$ for positive (negative) initial velocities of the time.\\
\noindent
$\bullet$  \textit{For constant forces} $F_0$ we have ${\cal V}(q) = - F_0 (q-q_0)$ and
we get the canonical equations of motion $t'=\mp\sqrt{m/\{2[{\cal H}+F_0(q-q_0)]\}}$ and
${\cal H}'=0$, leading to ${\cal H}={\cal H}_0$ and the time solution is
\begin{equation}
t(q)= t_0 \mp\frac{\sqrt{2m({\cal H}_0 +F_0(q-q_0))}}{F_0}.
\label{tcte}
\end{equation}
For the gravitational case $F_0=-mg$ we recover the solution
from Eq.~(\ref{tqfall}) when ${\cal H}_0=m/(2{t'_0}^2)-mgq_0$.  Assuming $q(0)=q_0=t(0)=t_0=0$, we have
\begin{equation}
t(q)= \pm\frac{\sqrt{2m({\cal H}_0 -mg q)}}{mg},
\label{tcte}
\end{equation}
and the velocity is
\begin{equation}
t'(q) =  \mp\frac{m}{\sqrt{2m({\cal H}_0 -mgq)}},
\end{equation}
which depends on the position. When $q$ increases from $q_0=0$ and reaches the turning point $q_{\mbox{\tiny tur}} = {\cal H}_0/mg$,  positive and negative times collapse to zero. The positive and negative velocities of the time diverge. When the particle starts to fall, the inverse situation occurs. Note that the velocity of the time is independent of $t_0$.

\noindent
$\bullet$ \textit{For the harmonic oscillator} (HO), the canonical equations of motion  are $t'=\mp\sqrt{m/[2({{\cal H}-kq^2/2})}]$ and ${\cal H}'=0$, which lead to the solutions  ${\cal H}={\cal H}_0$ and 
\begin{eqnarray}
t(q) &=& t_0\mp\frac{1}{\omega}\arctan{\frac{\sqrt{k}q}{\sqrt{2({\cal H}_0-kq^2/2)}}},\cr
& &  \cr
t(q) &=& t_0 \mp \frac{1}{\omega}\arccos{\left(\frac{q}{A}\right)},
\label{td}
\end{eqnarray} 
where we used the identity $\arctan{(x)} = \arccos{(1/\sqrt{1+x^2})}$, for $x>0$.  Here $\omega^2=k/m$ and $A=\sqrt{q_0^2+v_0^2/\omega^2}$. Equation (\ref{td}) is exactly the solution for the HO with  $t_0=-\phi/\omega$, with $\phi$ being a phase, and ${\cal H}_0= mv/(2{t'_0}^2)+kq_0^2/2 = m\omega^2A^2/2$ is identified as the total initial mechanical energy. Starting at $q=0$ with $t_0=0$, the negative(positive) time increases(decreases) and collapse to zero at the turning points. The positive and negative velocities of the time are
\begin{equation}
t'(q)= \pm\frac{1}{\omega\sqrt{A^2-q^2}},
\label{txHl}
\end{equation}
and also diverge at the turning points.

In all examples, inverting the solutions, we obtain the usual correct solutions $q(t)$ so that the classical dynamics are unaffected when considering the position as an independent variable, as expected. Furthermore, the positive and negative branches of the time, and the positive and negative velocities of the time,  do not affect the classical solutions.

\section{The quantum version}
\label{Qversion}

The common quantization procedure is to transform conjugate variables in operators. Therefore, we may use the Momentumian (\ref{PP}) to obtain the operator $\hat {\cal P}^{\pm}(\hat t,\hat {\cal H};q)$ and use it to describe the quantum dynamics. Surprisingly, such an operator is in full agreement with the operator $\hat {P}(\hat t,\hat H;q)$, proposed recently in a  spacetime symmetric formalism for the QM \cite{diasparisio} for time-independent potentials.   In their case, the complete Hilbert space is divided into  
$\mathcal{H} = \mathcal{H}_{\text{pos}} \otimes \mathcal{H}_{\text{time}},$
where $\mathcal{H}_{\text{pos}}$ is the usual Hilbert space and $\mathcal{H}_{\text{time}}$ is a temporal \emph{extension} of the regular theory. In the extended space
$\mathcal{H}_{\text{time}}$, we have the time operator $\hat{t}$ with eigenkets $\left|t \right>$ as
$\hat{t} \left| t \right> = t \left| t \right>$,
where $t$ is the eigenvalue associated to $\left|t \right>$. The set of eigenkets $\{\left| t \right>\}$ satisfy the identity $\mathbb{I} = \int_{-\infty}^{\infty} \mathrm{d} t \, \left| t \middle> \middle< t \right|$. The momentum operator proposed  \textit{ad hoc} in \cite{diasparisio} is $\hat {P}(\hat t,\hat H;q)=\pm\sqrt{2m[\hat H-V(\hat t;q)]}$, which has the same form as the Momentumian from Eq.~(\ref{PP}) when assuming that $t$ and ${\cal H}$ are operators. If we consider the operators $\hat t$ and $\hat {\cal H}$, the connection  between $\hat {P}(\hat t,\hat H;q)$ and $\hat {\cal P}^{\pm}(\hat t,\hat {\cal H};q)$ is evident.  

Call to mind the Schrödinger equation (SE) is given by $i\hbar \partial_t |\psi(t)\rangle= \hat H(\hat p,q;t)  |\psi(t)\rangle$ where  $|\psi(t)\rangle$ is the ket describing the quantum state in the Hilbert space, and $H(\hat p,\hat q;t)=T(\hat p)+V(\hat q)$ is the Hamilton operator which generates the time translation of  $|\psi(t)\rangle$. Here, the time $t$ is a parameter. However, in the spacetime symmetric description of Ref.~\cite{diasparisio},  the quantum state in the ${\cal  H}$ilbert space is defined as the ket $|\phi(q)\rangle$, which has its corresponding bra $\langle\phi(q)|$ satisfying the usual QM properties. The canonical variables ($t,-{\cal H}$) become operators ($\hat t,-\hat {\cal H}$), and the position $q$ is now the parameter.  The equation which generates the spatial translation of the ket $|\phi(q)\rangle$, as defined in Ref.~\cite{diasparisio} in analogy to the Schrödinger equation (check further also Refs.~\cite{diasparisioexp, arlans23-1, araujo2023} for more details), is 
\begin{equation}
\hat {\cal P}^{\pm}(\hat t,\hat {\cal H};q) |\phi(q)\rangle = - i\hbar\frac{\partial}{\partial q} |\phi(q)\rangle,
\label{SchP}
\end{equation}
with  the quantum Momentumian operator
\begin{equation}
{\cal \hat P}^{\pm}(\hat t,\hat {\cal H};q)=\pm \sqrt{2m({\hat {\cal H}} - {\cal V}(q))}.
\label{PKQ}
\end{equation}
We mention that the Momentumian is the quantum operator which generates the space translation of the ket. The solution for  Eq.~(\ref{SchP}) is $|\phi(q+dq)\rangle = {\cal U}^{\pm}|\phi(q)\rangle$ with 
\begin{equation}
{\cal U}^{\pm}(q,q_0)= \exp{\left[\frac{i}{\hbar} \int_{q_0}^q{\cal \hat P}^{\pm}(\hat t,{\cal \hat H};q) dq\right]},
\label{U}
\end{equation}
being the \textit{space} evolution operator, where we assumed a space-ordered operator, in analogy with the time evolution operator for time-dependent Hamiltonians in the usual QM. 

In this work, we focus and discuss the quantum properties solely related to the subspace $\mathcal{H}_{\text{time}}$ of the spacetime symmetric formalism, obtained through standard quantization techniques from Classical Mechanics.   Since this is an extended portion of the Hilbert space, we do not necessarily expect to get the same properties from $\mathcal{H}_{\text{pos}}$, but instead,  additional and/or complementary information about the quantum dynamics. For a complete description in the whole Hilbert space, we refer to \cite{diasparisio}.

For completeness, in the associated conjugated space to $\hat t$, we  have the  operator $\hat{\cal H}$ with eigenkets $\left|{\cal H}\right>$ as $\hat{\cal H} \left| {\cal H} \right> = {\cal H} \left| {\cal H} \right>$, where ${\cal H}$ is the eigenvalue associated to $\hat {\cal H}$.  The set $\{\left| {\cal H} \right>\}$ satisfies the identity $\mathbb{I} = \int_{-\infty}^{\infty} \mathrm{d} {\cal H} \, \left| {\cal H} \middle> \middle< {\cal H} \right|$. Wave functions in both representations are related through the Fourier transforms
\begin{eqnarray}
\phi(t;q) &=&\frac{1}{\sqrt{2\pi\hbar}}\int_{-\infty}^{\infty}\Phi({\cal H};q)e^{-i{\cal H}t/\hbar}d{\cal H},\\
& & \cr
\Phi({\cal H};q) &=&\frac{1}{\sqrt{2\pi\hbar}}\int_{-\infty}^{\infty}\phi(t;q)e^{+i{\cal H}t/\hbar}dt.
\end{eqnarray}
With this, it is straightforward to show that the operators act as $\hat t\,|{\cal H}\rangle= -i\hbar \partial_{\cal H}\,|{\cal H}\rangle $ and  $\hat {\cal H}\,|t\rangle=i\hbar \partial_{t}\,|t\rangle$. Besides, we have that $[\hat t,(-\hat {\cal H})]=i\hbar$, in agreement with the canonical quantization of the classical Poisson brackets.

\subsection{Separation of variables}

Next, we apply Dirac's procedure~\cite{dirac}, which in our case, allows the separation of variables in Eq.~(\ref{SchP}). For this we propose $ {\cal \hat P} = \alpha\sqrt{2m{\cal \hat H}} - \beta\sqrt{2m{\cal \hat V}(q)}$, where $\alpha$ and $\beta$ are $2 \times 2$ matrices, so that
\begin{eqnarray}
&\left({\cal \hat P} - \alpha\sqrt{2m{\cal \hat H}} + \beta\sqrt{2m{\cal \hat V}(q)} \right)\times\nonumber \\
&\left({\cal \hat P} + \alpha\sqrt{2m{\cal \hat H}} - \beta\sqrt{2m{\cal \hat V}(q)} \right)
\left| \phi(q) \right>=0,\nonumber
\end{eqnarray}
is compared to
\begin{eqnarray}
\left[{\cal \hat P}^2 - 2m({\cal \hat H} - {\cal \hat V}(q)) \right]\left| \phi(q) \right>=0.
\label{sqP}
\end{eqnarray}
The solution ${\cal \hat P}<0$ from Eq.~(\ref{sqP}) is also a physical solution of Eq.~(\ref{SchP}).  As a result, for $[\hat{\cal H}, \hat{\cal V}] = 0$ we find that
\begin{eqnarray*}\alpha=\left( \begin{tabular}{cc} 
                         $0$ & $1$\\
                         $1$ & $0$
\end{tabular}    \right),\quad \beta=\left( \begin{tabular}{cc} 
                         $-i$ & $0$\\
                         $0$ & $i$
  \end{tabular}    \right),
  \end{eqnarray*}
($\alpha\beta+\beta\alpha=0$)  
so that Eq.~(\ref{SchP}) becomes the Dirac equation:
\begin{eqnarray}
\label{dirac}
 \left(\alpha\sqrt{2m \hat{{\cal H}}} -\beta\sqrt{2m\hat{{\cal V}}(q)}\right)\left|\phi(q)\right>&=&
- i\hbar\,\frac{\partial\left|\phi(q)\right>}{\partial q}.\nonumber
\end{eqnarray}

Applying $\langle t|$ from the left, and using  $\left< t \middle| \sqrt{\hat{{\cal H}}}  \middle| \phi(q) \right> = \sqrt{i \hbar \partial_t} \left< t \middle| \phi(q) \right>$ we obtain, 
\begin{eqnarray}
 \left(\alpha\sqrt{2mi\hbar\partial_t} -\beta\sqrt{2m{\cal V}(q)}\right)\phi(t;q)&=&
 - i\hbar\,\partial_q\phi(t;q),\nonumber
\end{eqnarray}
where  $\phi(t|q)$  has two components $\phi^{\pm}(t|q)$, related to  ${\cal P}^{\pm}(\hat t,{\cal \hat H};q)$. Thus, we end up with a system of two coupled fractional Dirac equations:
\begin{eqnarray}
      \left\{ \begin{array}{ccc}
         \sqrt{2 m i\hbar\partial_t} \phi^{+}(t|q) &=& -\left(i \hbar  \partial_q + i\sqrt{2 m {\cal V}(q)}\right)  \phi^{-}(t|q),\\
     &&\cr
        \sqrt{2 m i\hbar\partial_t} \phi^{-}(t|q)  &=&  -\left(i \hbar  \partial_q- i\sqrt{2 m {\cal V}(q)}\right)\phi^{+}(t|q).
      \end{array} \right. \nonumber
  \end{eqnarray}
  {Using separation of variables $\phi^{\pm}(t;q)=\psi^{\pm}(t)\chi^{\pm}(q)$ allow us to rewrite the above equations in the following form:
    \begin{eqnarray}
    \label{classallowregionsepar}
      \left\{ \begin{array}{rcl}
        \chi^{+}(q)\sqrt{2 m i\hbar\,\partial_t} \psi^{+}(t) &=& - i \hbar   \psi^{-}(t)\partial_q \chi^{-}(q)\\[5pt]
        && - i \psi^{-}(t)\sqrt{2 m {\cal V}(q)}  \chi^{-}(q),\\
     &&\\
        \chi^{-}(q)\sqrt{2 m i\hbar\,\partial_t} \psi^{-}(t)  &=&  -i \hbar  \psi^{+}(t) \partial_q \chi^{+}(q)\\[5pt]
        &&+ i \psi^{+}(t)\sqrt{2 m {\cal V}(q)}\chi^{+}(q).
      \end{array} \right.
  \end{eqnarray}
In the following, we consider the TIDE and the PIDE separately.



\subsubsection{Position-independent equations}  
Applying the method of separation of variables to Eqs.~(\ref{classallowregionsepar}), we obtain two coupled linear \textit{position-independent} equations
\begin{eqnarray}
\sqrt{2mi\hbar}\,D_t^{1/2}\psi^+(t) &=& P_t^+\psi^-(t), \label{psi1}\\
&&\cr
\sqrt{2mi\hbar}\,D_t^{1/2}\psi^-(t) &=& P_t^-\psi^+(t), \label{psi2}
\end{eqnarray}
where $D_t^{1/2}$ is the $1/2$-fractional Caputo derivative, $P_t^+=m/t'_+$ and $P_t^-=m/t'_-$ are separation constants, and are the initial (at $q_0$) momentum in time. In the usual QM, the separation constant is the total energy $E$. Here we have two possible initial momenta, which may have distinct values and signs. Equations (\ref{psi1}) and (\ref{psi2}) are fractional differential equations whose solutions depend on the considered kind of fractional derivative. There are some proposals in the literature for fractional derivatives. However, for physical systems, it seems to be more appropriate to use the Caputo type \footnote{Since the Caputo derivative of a constant is zero, and the method of using Laplace transforms to solve differential equations involves only initial conditions on derivatives of integer order (which may not be the case depending on the type of fractional operators), we consider Caputo fractional derivative to be more appropriate for physical systems. }  with $\alpha=1/2$ \cite{tarasov}, which we adopt here. The solution for the position-independent equations is then given explicitly by \cite{lim12} (see also \cite{laskin,duan18})
\begin{eqnarray}
  \label{psiG}
  \left(\begin{array}{c}
    \psi^+(t) \\
    \psi^-(t)
  \end{array} \right) &=& A_0 E_{\frac{1}{2}}\left[\sqrt{\frac{P_t^+P_t^-t}{2m\,i\hbar}}\right] \left( \begin{array}{c}
    1 \\
    \sqrt{\frac{p_t^+}{p_t^-}}
  \end{array} \right) \nonumber \\
  && + B_0 E_{\frac{1}{2}}\left[-\sqrt{\frac{P_t^+P_t^-t}{2m\,i\hbar}}\right] \left( \begin{array}{c}
    1 \\
    -\sqrt{\frac{p_t^+}{p_t^-}}
    \end{array}
    \right)\nonumber\\
\end{eqnarray}
with $A_0= \psi^+(0) + {\sqrt{p_t^-}\psi^-(0)/\sqrt{p_t^+}}$, $B_0=\psi^+(0)- {\sqrt{p_t^-}\psi^-(0)/\sqrt{p_t^+}}$, and 
$E_{\alpha}[{z^{\alpha}}]$ is the  Mittag-Leffler function defined by
\begin{equation}
E_{\alpha}({z^{\alpha}}) =\sum_{k=0}^{\infty}\frac{{z^{\alpha k}}}{\Gamma(\alpha k+1)},
\end{equation}
with $z$ being a complex number and $\alpha>0$. For $\alpha=1$ it becomes
$E_{1}(z) =\exp{(z)}$, so one can understand the Mittag-Leffler function as a generalization of the exponential function. Solution (\ref{psiG}) assumes that the  initial derivatives are $ \frac{d}{dt}\psi^+(0)= \frac{d}{dt}\psi^-(0)=0$,  and can be rewritten using the identity $E_{1/2}[z^{1/2}] = \exp{[z]}\operatorname{erfc}{[-z^{1/2}]}$.
The constants $A_0$ and $B_0$ from solution Eq.~(\ref{psiG}) can be determined by using $\psi^+(0)=\psi^-(0)=1$. 
Note that from Eqs.~(\ref{psi1}) and (\ref{psi2}), for  $P_t^+=P_t^-=0$ both, $\psi^+(t)$ and $\psi^-(t)$, are constants and no time dependence is observed.

Figure \ref{fig2} displays $|\psi^+(t)|^2$ as a function of $t$, the eigenvalue of $|t\rangle$, for $A_0=1$ and $B_0=0$. It starts at $1$ and increases quickly, oscillating and converging to $4$ for large $t$. The oscillations become faster by increasing the frequency ${\cal K}_0/\hbar$. Here ${\cal K}_0=P_t^+P_t^-/2m$ is the kinetic energy of the time. 
\begin{figure}[!h]
	\centering
	\includegraphics[width=0.9\columnwidth]{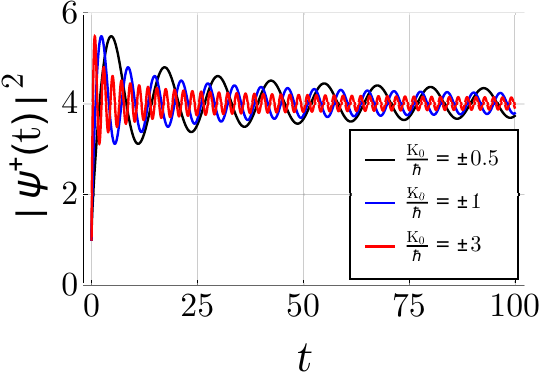}
	\caption{$|\psi^{+}(t;q)|^2$ as a function of the eigenvalue $t$, and fixed value of $q$ for distinct values of the frequency ${\cal K}_0/\hbar$.}
	\label{fig2}
\end{figure}

Due to Dirac's procedure, a new feature appears when compared to the usual position-independent SE solution, namely, we are now able to combine the separation constants $P_t^+$ and $P_t^-$: \\ 

\noindent
 \underline{\{F\}}: \textit{Forward} in time solutions for ${\cal K}_0>0$ ($P_t^+P_t^-=(m/2t'_+)(m/2t'_-)>0$).\\

 \noindent
 \underline{\{$\emptyset$\}}:   \textit{Absence} of time evolution for ${\cal K}_0=0$ ($P_t^+P_t^-=0$).  \\
 \\ 
  \noindent
\underline{\{B\}}:  \textit{Backward} in time evolutions for ${\cal K}_0<0$ ($P_t^+P_t^-<0$). \\

Crucial to realize that the time evolution is obtained using the kinetic energy  ${\cal K}_0$ (a constant), and not the total energy, as usual. Furthermore, ${\cal K}_0=0$ is the crossing point between forward and backward time evolutions.

\subsubsection{Time-independent equations}  
From Eq.~(\ref{classallowregionsepar}), the two first-order coupled linear \textit{time-independent} Dirac equations become 
\begin{eqnarray}
  -i\hbar \,\frac{d}{dq}\chi^-(q) -i\sqrt{2m{\cal V}(q)}\chi^-(q)&=& 
  P_t^+\chi^+(q),\label{chi1}\\
 &&\cr
  -i\hbar\,\frac{d}{dq}\chi^+(q) +i\sqrt{2m{\cal V}(q)}\chi^+(q)&=& 
  P_t^-\chi^-(q),\label{chi2}
\end{eqnarray}
which does not involve any fractional derivative. Equations (\ref{chi1}) and (\ref{chi2}) are the main result regarding time-independent equations and will be discussed next. To better compare our results to the usual QM approach, we isolate $\chi^+$ [$\chi^-$] in Eq.~(\ref{chi1}) [(\ref{chi2})] and substitute it in Eq.~(\ref{chi2}) [(\ref{chi1})], and obtain}
\begin{eqnarray}
\label{2N}
\left[-\frac{\hbar^2}{2m}\frac{d^2}{dq^2}+{\cal V}^{(\pm)}_{\mbox{\tiny eff}}(q)\right]\chi^{\pm}(q) ={\cal K}_0\chi^{\pm}(q),
\end{eqnarray}
with the effective potential 
\begin{eqnarray}
\label{Veff1}
{\cal V}^{(\pm)}_{\mbox{\tiny eff}}(q) &=& {\cal V}(q)\pm\frac{\hbar}{\sqrt{2m}}\frac{d}{dq}\left(\sqrt{{\cal V}(q)}\right),\\
&=& {\cal V}(q)\mp \frac{\hbar}{2\sqrt{2m{\cal V}(q)}}{\cal F}(q),
\end{eqnarray}
where  ${\cal F}(q)=- d\,{\cal V}(q)/dq$ is the force and $\sqrt{2m{\cal V}(q)}$ is the characteristic momentum of the potential. Equation (\ref{2N}) is the usual TISE for the effective potential ${\cal V}^{(\pm)}_{\mbox{\tiny eff}}(q)$ and the energy ${\cal K}_0$. Let us discuss separately the coupled and uncoupled cases.\\

\textit{Uncoupled equations.}  For the separable case $P_t^+=P_t^-=0$ (${\cal K}_0=0$, no time evolution) the solutions for Eqs.~(\ref{chi1}) and (\ref{chi2}) are
\begin{equation}
\chi_{\emptyset}^{\pm}(q)=\chi_{\emptyset}^{\pm}(0)\,e^{\pm\frac{\sqrt{2m}}{\hbar}{\int_0^q}\sqrt{{\cal V}(\overline q)}d\overline q}.
\label{V0}
\end{equation}
\begin{figure}[!h]
	\centering
	\includegraphics[width=1.1\columnwidth]{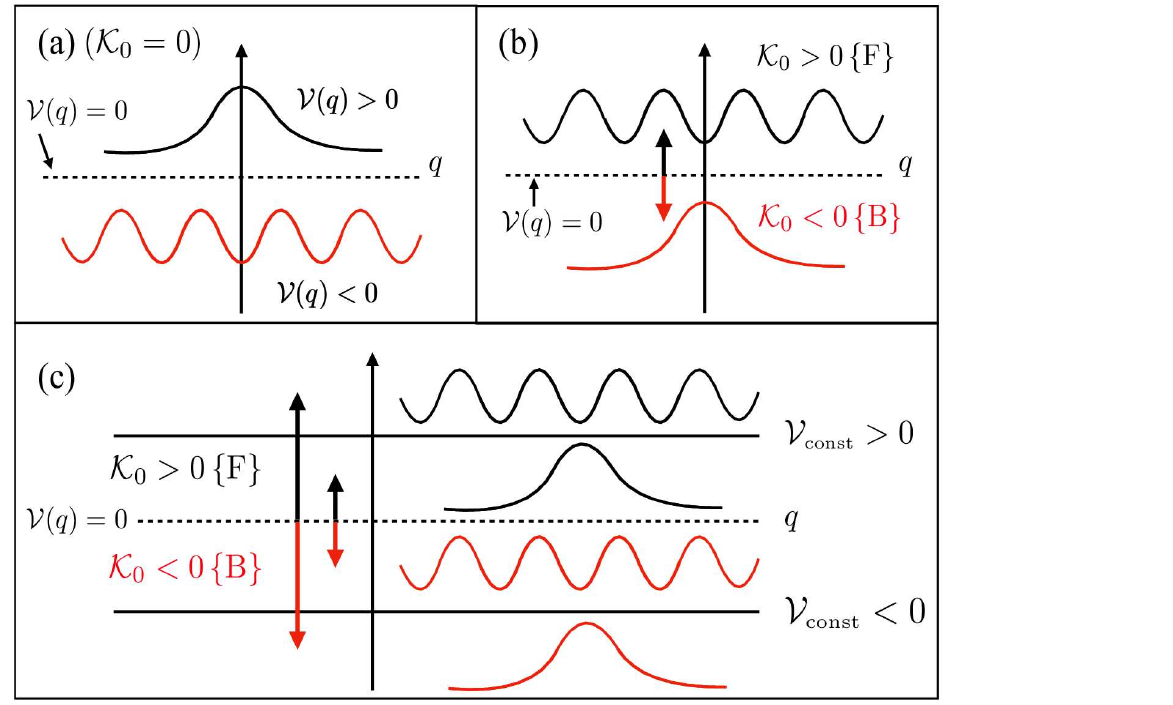}
	\caption{Schematic representation of the approximated wave functions behaviour in distinct situations: (a) Decoupled Eqs.(\ref{chi1}) and (\ref{chi2}) for ${\cal K}_0=0$, no time evolution. (b) Coupled case for  ${\cal V}(q)=0$ and kinetic energy ${\cal K}_0\ne 0$. The black (red) arrow represents the amount of kinetic positive (negative) energy inserted into the system, leading to forward (backwards) in-time solutions. (c) Coupled case with two exemplary constant potential energies, namely ${\cal V}_{\mbox{\tiny const}}>0$ and ${\cal V}_{\mbox{\tiny const}}<0$. At the dashed lines we have  ${\cal V}(q)= {\cal K}_0=0$.}
	\label{Emn}
\end{figure}
The wave functions are constants in the case of ${\cal V}(q)=0$. For ${\cal V}(q)<0$, we have oscillatory solutions, and for ${\cal V}(q)>0$, we have decaying and(or)  diverging solutions, depending on the form of ${\cal V}(q)$. This is schematically shown in Fig.~\ref{Emn}(a) for the non-diverging solutions. There is a clear distinction between the wave functions for positive and negative potential energies. Equation (\ref{2N}) can be written as an effective Hamiltonian satisfying ${\cal H}^{(\pm)}_{\mbox{\tiny eff}}\chi_{\emptyset}^{\pm}(q)=0$, with zero energy.  Alternatively, we can put the extra term in the potential on the right-hand side and interpret it as effective energy. This becomes clear in the examples discussed later on.

\textit{Coupled equations.} In this case ${\cal K}_0\ne0$ and the wave functions $\chi^{\pm}(q)$ are coupled via Eqs.~(\ref{chi1}) and  (\ref{chi2}).
{We start with ${\cal V}(q)={\cal V}(q_0)=0$, so that the time-independent Eq.~(\ref{2N}) becomes 
 \begin{eqnarray}
\label{2N0}
-\frac{\hbar^2}{2m}\frac{d^2}{dq^2}\chi^{\pm}(q)={\cal K}_0\chi^{\pm}(q),
\nonumber
\end{eqnarray}
which is a HO equation for the wave functions $\chi^{\pm}(q)$ with frequency $\omega=\sqrt{2m{\cal K}_0}/\hbar$. In other words, for  $\chi_{\mbox{\tiny F}}^{+}= \chi_{\mbox{\tiny F}}^{-}=\chi_{\mbox{\tiny F}}$ we have the oscillatory solutions 
    \begin{eqnarray}
\chi_{\mbox{\tiny F}}(q)  &=& \left[C_1e^{i\omega_{\mbox{\tiny F}}q}+ C_2 e^{-i\omega_{\mbox{\tiny F}}q} \right], \label{x1}
    \end{eqnarray}
with frequency $\omega_{\mbox{\tiny F}}=\,P_t/\hbar$ for case \{F\} and an inverted HO with  $\omega_{\mbox{\tiny B}}=i\,P_t/\hbar$ for case \{B\}.  Note that only for case \{F\} the above solutions are the expected plane wave solutions for the free particle. Otherwise, for case \{B\}, solutions will diverge/vanish with increasing positive/negative position $q$. See also Fig.~\ref{Emn}(b). The dynamics here is also distinct for positive or negative  ${\cal K}_0$, and the solutions $\chi^{\pm}(q)$ are \textit{coupled} via Eqs.~(\ref{chi1}) and (\ref{chi2}).

Next, we assume ${\cal V}(q) = {\cal V}_{\mbox{\tiny const}}$ being a constant, and the solutions (\ref{psiG}) are still valid for the time evolution. The time-independent Eq.~(\ref{2N}) becomes 
 \begin{equation}
\label{2NV0}
\left(-\frac{\hbar^2}{2m}\frac{d^2}{dq^2}+{\cal V}_{\mbox{\tiny const}}\right)\chi^{\pm}(q)=
{\cal K}_0\chi^{\pm}(q),
\end{equation}
which is the HO equation for the wave functions $\chi^{\pm}(q)$ with frequency $\omega^{\mbox{\tiny (const)}}_{\mbox{\tiny F}}=\sqrt{2m({\cal K}_0-{\cal V}_{\mbox{\tiny const}})}/\hbar$ for case \{F\}.  Solution (\ref{x1}) for \{F\} is still valid by using $\omega_{\mbox{\tiny F}}=\omega^{\mbox{\tiny (const)}}_{\mbox{\tiny F}}$. For ${\cal K}_0>{\cal V}_{\mbox{\tiny const}}$, oscillating functions are obtained [see left black arrow in Fig.~\ref{Emn}(c)]. For kinetic energies bellow the potential, ${\cal K}_0<{\cal V}_{\mbox{\tiny const}}$, $\omega^{\mbox{\tiny (const)}}_{\mbox{\tiny F}}=i\sqrt{2m(|{\cal K}_0-{\cal V}_{\mbox{\tiny const}}|)}/\hbar=i\,P_t/\hbar$ becomes imaginary and the solution (\ref{x1}) decays exponentially with $q$ [see right black arrow in Fig.~\ref{Emn}(c)]. For case \{B\}, we also obtain  $\omega^{\mbox{\tiny (const)}}_{\mbox{\tiny B}}=i\,P_t/\hbar$. 
Figure \ref{Emn}(c) summarizes the effect of  ${\cal K}_0$ when positive and negative constant potentials ${\cal V}_{\mbox{\tiny const}}$ are present.
For ${\cal V}_{\mbox{\tiny const}}>0$, the decaying solution is only transformed into an oscillating one when ${\cal K}_0> {\cal V}_{\mbox{\tiny const}}$ (see black arrows). For ${\cal V}_{\mbox{\tiny const}}<0$, the oscillating solution is only transformed into a decaying solution when $-{\cal K}_0< -{\cal V}_{\mbox{\tiny const}}$ (see red arrows).

\subsection{Applications for $F(q)\ne 0$}

\subsubsection{Harmonic potential.} 
\label{SHO}
In this case, the effective potential from (\ref{Veff1}) reduces to
\begin{eqnarray}
\label{VeffHO}
{\cal V}^{(\pm)}_{\mbox{\tiny eff}}(q)&=&\frac{kq^2}{2}\pm \frac{\hbar\omega}{2},
\end{eqnarray}
with $\omega=\sqrt{k/m}$. Equation~(\ref{2N}) can be rewritten as (the additional term $\mp\hbar\omega/2$ is inserted in the total energy on the right)
\begin{eqnarray}
\label{2NHO}
\left(-\frac{\hbar^2}{2m}\frac{d^2}{dq^2}+\frac{m\omega^2}{2}q^2 \right)\chi^{\pm}={\cal E}^{\pm}_{\mbox{\tiny eff}}
\chi^{\pm},
\end{eqnarray}
which is exactly the TISE whose quantized energy ${\cal E}_n$ is well known and given by 
\begin{equation}
\label{nQ}
{\cal E}_n= \left(n+\frac{1}{2}\right)\hbar\omega = {\cal K}_0\mp \frac{\hbar\omega}{2} ={\cal E}^{\pm}_{\mbox{\tiny eff}},\, n=0,1,2,\ldots.
\end{equation}
Thus, our interpretation does not change the energy quantization in the HO. However, this leads to ${\cal K}_0$ quantization. ${\cal K}_0$ is a constant introduced via separation constants $P_t^{\pm}$. It does not mean that the kinetic energy is constant but that the energy inserted via the separation constants must be quantized to obtain the correct quantization of the HO. To better understand the underlying physics, we have to go back to Eqs.~(\ref{chi1}) and (\ref{chi2}) and consider again the two situations.

\textit{Decoupled equations}. For the separable case $P_t^+=P_t^-=0$, or ${\cal K}_0=0$ (no time evolution), the solutions are
\begin{eqnarray}
\label{decSol}
\chi_{\emptyset}^-(q) = \left(\frac{m\omega}{\pi\hbar}\right)^{\frac{1}{4}}e^{-\frac{m\omega}{2\hbar}q^2},\quad \chi_{\emptyset}^+(q) = A e^{\frac{m\omega}{2\hbar}q^2},
\end{eqnarray}
with $A$ being a constant. While $\chi_{\emptyset}^-(q)$ is the exact solution of the HO for the ground state with energy $\hbar\omega/2$, $\chi_{\emptyset}^+(q)$ diverges for large $q$ values so that we must choose $A=0$. However, we mention that changing $q\to I \,q$,  $\chi_{\emptyset}^+(q)\to \chi_{\emptyset}^-(q)$ does not diverge anymore and becomes the eigenfunction of the ground state for the inverted harmonic potential. 

Nevertheless, deriving  Eqs.~(\ref{chi1}) and (\ref{chi2})  relative to $q$, we obtain the separated equations 
\begin{eqnarray}
\left(-\frac{\hbar^2}{2m}\frac{d^2}{dq^2}+\frac{m\omega^2q^2}{2}\right)\chi_{\emptyset}^{\pm}&=&\mp\frac{\hbar\omega}{2}\chi_{\emptyset}^{\pm}.\label{chi12}
\end{eqnarray}
Both equations are the TISE for the HO with energies given by the right-hand side. In other words, $\chi_{\emptyset}^-(q)$ is the solution for the equation of the ground state $n=0$, since the right-hand side of Eq.~(\ref{chi12}) is related to the energy $\hbar\omega/2$. On the other hand, $\chi_{\emptyset}^+(q)$ is the solution for the state equation for a negative energy $-\hbar\omega/2$, whatever it means, for the moment. Thus, \textit{if we do not add kinetic energy into the system (keeping ${\cal K}_0=0$), we naturally have a pair of uncoupled states (with energies $\pm\hbar\omega/2$) with no time evolution}. In other words, the ground state is stable. 
\begin{figure}[!h]
	\centering
	\includegraphics[width=1.0\columnwidth]{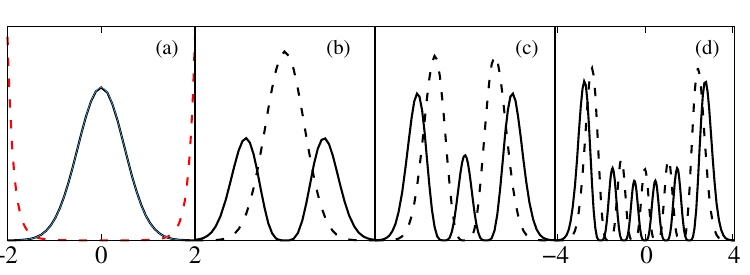}
	\caption{Normalized (numerically)  functions $|\chi^-(q)|^2$ (continuous curves) and $|\chi^+(q)|^2$ (dashed curves) for (a)  ${\cal K}_0=0$ (decoupled case), (b) ${\cal K}_0=\hbar\omega$, (c) ${\cal K}_0=2\,\hbar\omega$ and (d) ${\cal K}_0=5\,\hbar\omega$. The horizontal axis interval increases from $[-2.0,2.0]$ in panel (a) to $[-4.1,4.1]$ in panel (d).}
	\label{fig3}
\end{figure} 

Worthy of mentioning is that Eq.~(\ref{chi12}) for  $\chi_{\emptyset}^+(q)$ and $q\to i\,q$ (\textit{i.e.}, the inverted HO), is somehow related to the Bohmian optical potential responsible for repulsion of Bohmian trajectories in a focusing problem of free wave packets [see Eq.~(39) of \cite{briggsBohm}].

No excited states can be obtained as long ${\cal K}_0=0$. Figure \ref{fig3}(a) displays the expectation values $|\chi_{\emptyset}^{\pm}(q)|^2$ obtained by solving numerically Eqs.~(\ref{chi1}) and (\ref{chi2}). The normalization was done numerically inside the integrated interval of $q$. Ground state wave function for $\chi_{\emptyset}^-(q)$ (continuous black curve) and diverging solution for $\chi_{\emptyset}^+(q)$ (red dashed curve). Exact solution $\chi_{\emptyset}^-(q) $ from Eq.~(\ref{decSol}) is also plotted and completely matches with the numerical solution.

\textit{Coupled equations.}
For  ${\cal K}_0\ne 0$, states $\chi^+(q)$ and $\chi^-(q)$ become coupled via Eqs.~(\ref{chi1}) and (\ref{chi2}) and higher excited states are reached depending on the values of ${\cal K}_0$. Using the quantization of ${\cal K}_0$ from Eq.~(\ref{nQ}), we expect that only integer values of ${\cal K}_0/\hbar\omega$ lead to the correct eigenfunctions from the HO. This is indeed the case, as obtained by the numerical integration of Eqs.~(\ref{chi1}) and (\ref{chi2}), whose results can be observed by $|\chi_{\mbox{\tiny F}}^{(\pm)}(q)|^2$ in  Fig.~\ref{fig3}(b) for ${\cal K}_0=\hbar\omega$, Fig.~\ref{fig3}(c) for ${\cal K}_0=2\hbar\omega$, and  Fig.~\ref{fig3}(d) ${\cal K}_0=5\hbar\omega$. The wave function  $\chi_{\mbox{\tiny F}}^+(q)$ is always related to one state with lower energy than $\chi_{\mbox{\tiny F}}^-(q)$. Worthy to mention is that the constants $P_t^{(\pm)}$ in Eqs.~(\ref{chi1}) and (\ref{chi2})  need  not to be equal, but just satisfy ${\cal K}_0=P_t^+P_t^-/2m$. Thus, for a given value of $ {\cal K}_0$, we can change the values of  $P_t^{(\pm)}$ to control the probability of the states individually.

Above positive quantized energies, ${\cal E}_n= \left(n+\frac{1}{2}\right)\hbar\omega$, and their wave functions were obtained using the quantized values of ${\cal K}_0>0$ and contain forward in time evolutions (case \{A\}). Using the same procedure, but inserting negative quantized energies ${\cal K}_0<0$, the coupled Eqs.~(\ref{chi1}) and (\ref{chi2}) diverge and should be forced to be zero. However, using $q\to i\,q$ in Eqs.~(\ref{chi1}) and (\ref{chi2}), the wave functions from the HO are correctly reproduced with the corresponding negative spectrum ${\cal E}_n= -\left(n+\frac{1}{2}\right)\hbar\omega$, with $n>0$. This contains backward in time evolutions ${\cal K}_0<0$ (case \{B\}). However, to be consistent with the usual QM, since negative energies are not allowed in the usual QM of the HO, the case  ${\cal K}_0<0$ becomes prohibited. Nevertheless, the stationary negative ground state for ${\cal K}_0=0$ is present.

\begin{figure}[!h]
	\centering
	\includegraphics[width=1.0\columnwidth]{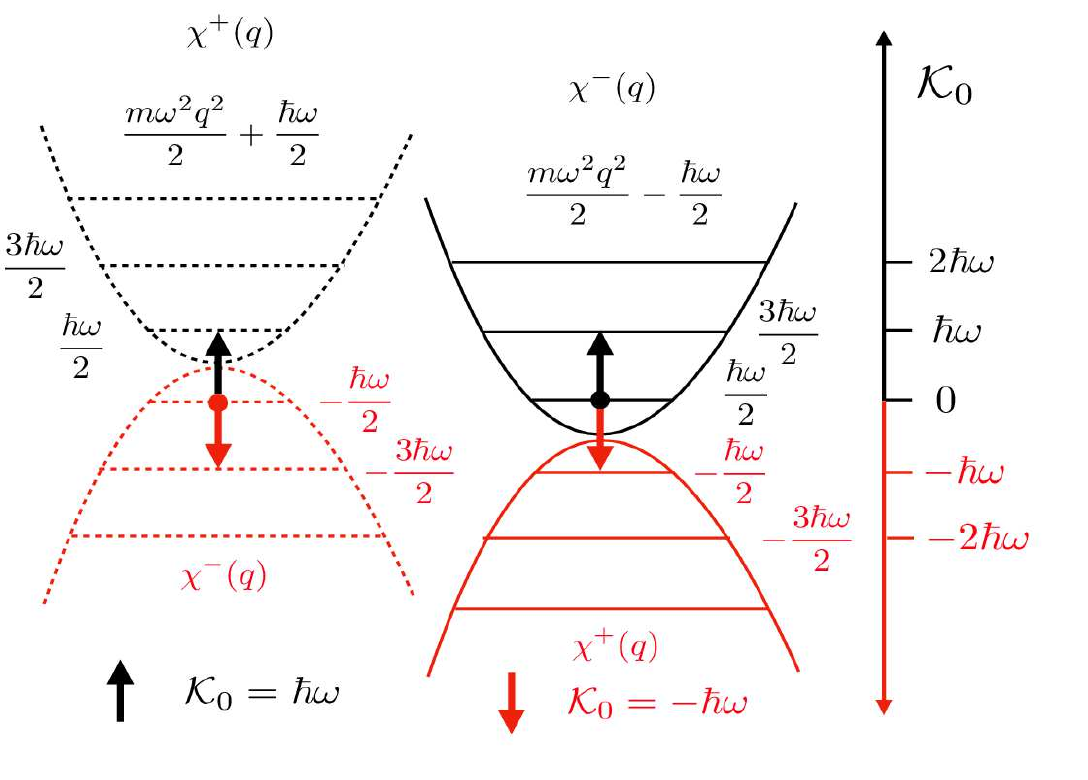}
	\caption{Schematic representation of the shifted harmonic potential  ${\cal V}^{(+)}_{\mbox{\tiny eff}}(q)$(left dashed curve) and  ${\cal V}^{(-)}_{\mbox{\tiny eff}}(q)$(right continuous curve). Red curves for the shifted inverted potentials. For the uncoupled case, ${\cal K}_0=0.0$, we have two solution, $\chi^-_0(q)$ for the ground state ($\bullet$) and  $\chi^+_0(q)$ for the ground state (\cbeims{$\bullet$}) in the inverted HO.  As an example, initial energy ${\cal K}_0=\hbar\omega$ induce transition upwards (black arrows), and  ${\cal K}_0=-\hbar\omega$ induce transition downwards (red arrows).}
	\label{fig4}
\end{figure} 
An alternative way to interpret the results is to look at the effective potential from Eq.~(\ref{VeffHO}). Figure \ref{fig4} displays schematically both shifted harmonic potentials in black. On the left we have ${\cal V}^{(+)}_{\mbox{\tiny eff}}(q)$ and on the right ${\cal V}^{(-)}_{\mbox{\tiny eff}}(q)$. The shifted potentials in red are the inverted HO potentials obtained by using $q\to iq$. For the uncoupled case, ${\cal K}_0=0.0$, we have two solutions, $\chi^+_{\emptyset}(q)$ for the ground state ($\bullet$) and  $\chi^-_{\emptyset}(q)$ for the ground state (\cbeims{$\bullet$}) in the inverted HO (if it exists). By applying forward-in-time evolution (say ${\cal K}_0=\hbar\omega$),  we induce transition upwards (black arrows). Applying backwards in time evolution (${\cal K}_0=-\hbar\omega$), we induce transition downwards (red arrows) using $q\to I \,q$.

Results are similar to Dirac's interpretation of the existence of antiparticles. For example, suppose the red state on the left of Fig.~\ref{fig4} is excited to the black state after receiving ${\cal K}_0=\hbar\omega$, as shown in the upward black arrow's illustration. Thus, we create a ``real particle'' inside the black potential on the left and a hole (the ``antiparticle'')  in the red potential on the left. Furthermore, the above example can only occur when the red state is coupled to the ground state on the right black potential, which necessarily must receive the same energy  ${\cal K}_0=\hbar\omega$. Thus, we can only create a particle on the left in Fig.~\ref{fig4} using the excitation of the black particle on the right. However, different from Dirac's result, here $\chi^+(q)\ne [\chi^-(q)]^*$.}

It can be argued that Eq.~(\ref{2NHO}) represents the same physics as in the usual QM with shifted potentials, 
\begin{eqnarray}
\label{2NHOusual}
\left(-\frac{\hbar^2}{2m}\frac{d^2}{dq^2}+\frac{m\omega^2}{2}q^2 \mp \frac{\hbar\omega}{2}\right)\tilde\chi^{\pm}(q)=E \tilde\chi^{\pm}(q),
\end{eqnarray}
with $E$ being the total energy, and $\tilde\chi^{\pm}(q)$ the solutions for the TISE with the shifts $\pm\hbar\omega/2$.  The important distinction between both approaches is that $\tilde\chi^{\pm}(q)$ above are uncoupled while $\chi^{\pm}(q)$ are coupled via Eqs.~(\ref{chi1}) and (\ref{chi2}). In other words, while Eq.~(\ref{2NHOusual}) represents two \textit{independent} shifted HOs, Eq.~(\ref{2NHO}) describes one HO behaving like two \textit{coupled} shifted HOs. In fact, in the latter, we have \textit{always} a pair of coupled states separated energetically by $\hbar\omega$. 

The above distinction becomes evident when we study the superposition of, for example, the ground and first excited states. In the usual MQ description, we have
\begin{eqnarray}
\Psi(q,t) &\propto&  \chi_{\emptyset_0}^-(q) e^{-i\frac{\omega}{2}t}+ \chi_{\emptyset_1}^-(q)e^{-i\frac{3\omega}{2}t},
\end{eqnarray}
with $\chi_{\emptyset_0}^-(q)$ for the ground state according to Eq.~(\ref{decSol}) and 
\begin{equation}
\chi_{\emptyset_1}^-(q)=\frac{\sqrt{2}}{\pi^{\frac{1}{4}}}\left( \frac{m\omega}{\hbar} \right)^{\frac{3}{4}}qe^{-\frac{m\omega}{2\hbar}q^2},
\nonumber
\end{equation}
for the first excited state. The time dependence is $|\Psi(q,t)|^2\propto \cos{\omega t}$. However, in the present approach, we have
\begin{figure}[!h]
	\centering
	\includegraphics[width=0.8\columnwidth]{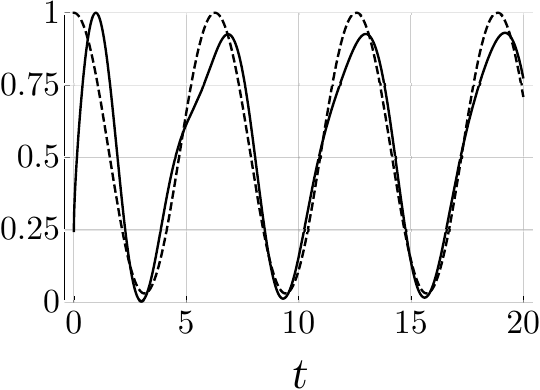}
	\caption{Plotted are $|\Psi(q;t)|^2$ (dashed curve) and   $|\phi(t;q)|^2$ (continuous curve) as a function of the time for the superposition of first and second excited states of the HO, scaled to the maximum values of the absolute values.}
	\label{fig5}
\end{figure} 
\begin{eqnarray}
\phi(t;q) &\propto&  \chi_{\emptyset_0}^-(q) \mbox{E}_{\frac{1}{2}}\left[ \sqrt{-i\omega t} \right] + \chi_{\emptyset_1}^-(q)\mbox{E}_{\frac{1}{2}}\left[ \sqrt{-2 i\omega t} \right],
\nonumber
\end{eqnarray}
which leads to a time dependence of $|\phi(t;q)|^2$ shown in Fig.~\ref{fig5} (continuous curve). The time dependence comparison between both cases shows some distinctions due to the Mittag-Leffler function.

\subsubsection{One-dimensional Hydrogen Atom} 
One electron moving in the one-dimensional potential ${\cal V}(q)=-\frac{e^2}{|q|}$ is an old problem, and the quantized energies are given by \cite{loudon59,loudon16} 
\begin{equation}
\label{En1D}
E_n=-\frac{\hbar^2}{2ma_0^2}\frac{1}{n^2},\quad n=1,2,3,\ldots,
\end{equation}
where $a_0=\hbar^2/(me^2)$ is the Bohr radius. The normalized wave functions, which are solutions of the TISE, are 
$\sqrt{2/(a_0^2n^5(n!)^2)}\exp{\left(-\frac{|q|}{n a_0}\right)}q L_n\left(\frac{2|q|}{n a_0}\right)$  for odd states and $\sqrt{2/(a_0^2n^5(n!)^2)}\exp{\left(-\frac{|q|}{n a_0}\right)}|q|L_n\left(\frac{2|q|}{n a_0}\right)$ for even states \cite{loudon59}.
$L_n(z)$ are the generalized Laguerre functions. 
The effective potential becomes 
\begin{eqnarray}
\label{VeffHy}
{\cal V}^{\pm}_{\mbox{\tiny eff}}(q)&=&-\frac{e^2}{|q|}\pm {\frac{i\,e\hbar}{2\sqrt{2m}}}\frac{1}{|q|^\frac{3}{2}},\cr
& & \cr
&=& -\frac{e^2}{|q|}\pm\frac{\hbar^2}{2ma_0^2}\left(\frac{i}{\sqrt{2}}\frac{a_0^{\frac{3}{2}}}{|q|^{\frac{3}{2}}}\right),
\end{eqnarray}
and the shift in the potential is imaginary and depends on $|q|$. 

\textit{Uncoupled case.} Using $P_t^+=P_t^-=0$ (${\cal K}_0=0$), the solutions  for Eqs.~(\ref{chi1}) and (\ref{chi2}) are
\begin{equation}
\label{chi1D}
\chi_{\emptyset}^{\pm}(q)=A_{\emptyset}^{\pm} e^{\pm\frac{i}{\hbar}\sqrt{8me^2|q|}},
\end{equation}
with $A_0^{\pm}$ being normalization constants. These are oscillating functions, solutions of ${\cal H}_{\mbox{\tiny eff}}\chi^{\pm}= 0$ and,  as for the HO, seem to be solutions around the zero potential energy, which in this case is the threshold between bounded and continuum states. This becomes clear in the numerical simulations for the coupled case discussed below. The term $\pm \sqrt{8me^2|q|}$ in the exponent has units of momentum $\times$ position. Thus, $\chi_{\emptyset}^{\pm}(q)$ can be interpreted as plane wave solutions. 
The difference to the usual plane waves is that for increasing (decreasing) $|q|$ values, the wavelength of such plane waves increases (decrease). Furthermore, we call to attention that by changing $e\to-e$, we interchange the solutions $\chi^+\leftrightarrow\chi^-$.

\begin{figure}[!h]
	\centering
	\includegraphics[width=1.0\columnwidth]{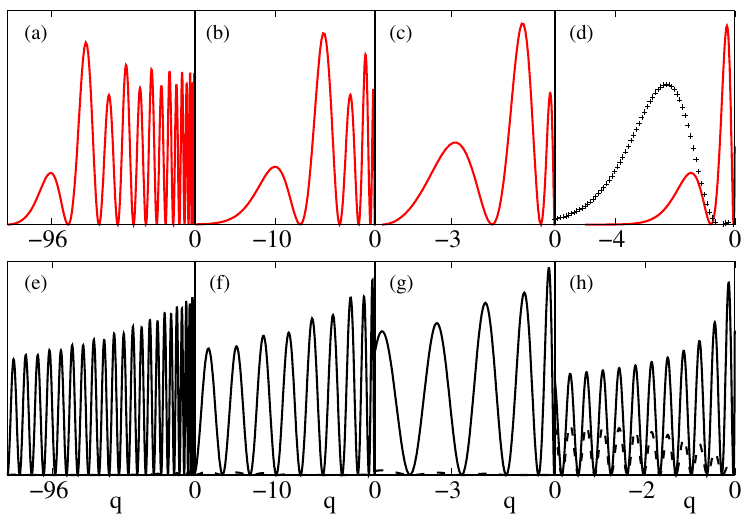}
	\caption{Normalized (numerically)  functions $|\chi^-(q)|^2$ (continuous curves) and $|\chi^+(q)|^2$ (dashed curves): (a) for ${\cal K}_0=-0.02$, (b) for ${\cal K}_0=-0.18$, (c)  for ${\cal K}_0=-0.5$ and (d)  for ${\cal K}_0=-1,28$. Panels (e)-(h) correspond to the same values of ${\cal K}_0$,  but positive. Red (black) for backwards (forward) in time evolutions.}
	\label{fig6}
\end{figure}

\textit{Coupled equations.} To study this case, we rely on the numerical integration of Eqs.~(\ref{chi1}) and (\ref{chi2}). We start with small values of the coupling and increase it. Namely, we choose the examples ${\cal K}_0=\pm 0.02, \pm 0.18, \pm 0.5$ and $\pm 1.28$. The normalized (numerically)  functions $|\chi^{\pm}(q)|^2$ are shown in Figs.~\ref{fig6} for $q<0$. Upper panels for ${\cal K}_0<0$ and bottom panels for  ${\cal K}_0>0$.  For ${\cal K}_0>0$ the solutions $|\chi_{\mbox{\tiny F}}^{\pm}(q)|^2$ are distinct, while for ${\cal K}_0<0$,  $|\chi_{\mbox{\tiny B}}^{\pm}(q)|^2$ are equal as long $P_t^+=P_t^-$. In this case, $\mbox{Re}[\chi_{\mbox{\tiny B}}^{+}(q)]=\mbox{Im}[\chi_{\mbox{\tiny B}}^{-}(q)]$ and  $\mbox{Im}[\chi_{\mbox{\tiny B}}^{+}(q)]=\mbox{Re}[\chi_{\mbox{\tiny B}}^{-}(q)]$. For $P_t^+\ne P_t^-$ the probabilities $|\chi_{\mbox{\tiny B}}^{\pm}(q)|^2$ become distinct (not shown).

Figure \ref{fig6}(a) shows that for small negative energy ${\cal K}_0=-0.02$ (bellow the threshold), $|\chi_{\mbox{\tiny B}}^{\pm}(q)|^2$ have many oscillations but converge to zero around $q=-126$, and should be related to a wave-function of an excited state. The corresponding Fig.~\ref{fig6}(e) for a small positive energy ${\cal K}_0=0.02$ (above the threshold), $|\chi_{\mbox{\tiny F}}^{+}(q)|^2$ is almost zero (not visible in the panel), but $|\chi_{\mbox{\tiny F}}^{-}(q)|^2$ oscillates continuously for $q\to-\infty$, as expected for a free particle.
Increasing the negative values of  ${\cal K}_0<0$, we go deeper into the negative energies of the atom and become wave functions which seem to be associated with lower energy states. For example, in Fig.~\ref{fig6}(d), the probability using the exact wave-function (cross black points) of the ground state $n=1$ is compared to $|\chi_{\mbox{\tiny B}}^-(q)|^2$ (red curve). The very small peak of the exact solution for small $q$-values, is enhanced in  $|\chi_{\mbox{\tiny B}}^-(q)|^2$. Both probabilities are not expected to be equal since one ${\cal K}_0$ value cannot compensate for the $q$-dependent of the shift. The wave-functions for the continuum in Fig.~\ref{fig6}(f)-(h) oscillate, even for $q\to-\infty$.

\vspace*{0.5cm}
\section{Conclusions}
\label{conclusions}

We want to summarize our main results. Using position as an independent parameter and time as a dependent variable, we obtain two branches of time in the classical solutions: one for increasing times and the other for decreasing times. Both branches lead to the same classical trajectories. The time ($t$), and minus the total energy ($-{\cal H}$), are the conjugated variables and the canonical equations of motion are obtained from the derived Momentumian ${\cal P}^{(\pm)}=\pm\sqrt{2m({\cal H}-{\cal V})}$, responsible for generating the canonical \textit{space} translation. The sign $(+)$ occurs when space and time have opposite directions, meaning that when space increases (decreases), time decreases (increases). Space and time have the same directions for the  $(-)$ sign. While classical dynamics remains unaltered, quantum dynamics show some different aspects. Firstly, the corresponding quantum operator   ${\cal \hat P}^{(\pm)}=\pm\sqrt{2m({\cal \hat H}-{\cal \hat V})}$ involves $1/2-$time fractional time derivatives.  As far we know, this is the first time fractional derivatives naturally appear from first principles, and ${\cal \hat P}^{(\pm)}$ is in full agreement with the operator proposed in a space-time-symmetric formalism of QM \cite{diasparisio}. Secondly, separation of variables in the quantum equation for ${\cal \hat P}^{(\pm)}$ is only possible using Dirac's procedure \cite{dirac} by proposing that $ {\cal \hat P} = \alpha\sqrt{2m{\cal \hat H}} - \beta\sqrt{2m{\cal \hat V}(q)}$, where $\alpha$ and $\beta$ are $2\times 2$ matrices. In this procedure we obtain two coupled $1/2$-order-position-independent Dirac equations (PIDEs), Eqs.~(\ref{psi1})-(\ref{psi2}), and two coupled first-order-time-independent Dirac equations (TIDEs), Eqs.~(\ref{chi1})-(\ref{chi2}). The effect of the separation of constants, namely the momenta $P_t^+$ and $P_t^-$, which always appear as a product, are quantified in ${\cal K}_0=P_t^+P_t^-/2m$, which is the kinetic energy (constant). Thus,  ${\cal K}_0$ quantifies the coupling between the equations and can be negative when $P_t^+$ and $P_t^+$ have opposite signals. For ${\cal K}_0=0$, the two PIDEs (and the two TIDEs) are decoupled, and no time evolution occurs.   PIDE can be solved using the Mittag-Lefller function, and the time evolution depends on ${\cal K}_0$. For ${\cal K}_0>0$, we obtain forward-in-time evolution and backwards-in-time evolution otherwise. The solutions of the TIDEs, $\chi^{\pm}(q)$, depend on the potential ${\cal V}(q)$. From both TIDEs, we can derive the TISEs, which contain positive/negative shifts in the potential, namely  $$\pm\frac{\hbar}{\sqrt{2m}}\frac{d}{dq}\left(\sqrt{{\cal V}(q)}\right)=\mp \frac{\hbar}{2\sqrt{2m{\cal V}(q)}}{\cal F}(q),$$ which are proportional to the force on the particle. A positive shift for  $\chi^{+}(q)$, and a negative one for $\chi^{-}(q)$.  In the case of zero forces, we obtain that  $\chi^{+}(q)=\chi^{-}(q)$. For finite forces, we discuss the HO and the one-dimensional Hydrogen atom.

In the HO case, the shifts in the potential are $\pm\hbar\omega/2$. For the decoupled case ${\cal K}_0=0$, we have two stable (no time evolution) ground states with energies $\pm\hbar\omega/2$. For ${\cal K}_0>0$, the quantization of the HO energies $(n+1/2)\hbar\omega$ leads to the quantization of ${\cal K}_0=k\hbar\omega$, with $k=1,2,\ldots$. For $k=1$, the coupled pair of states with quantum numbers $n=1,2$ are obtained for the HO. For  $k=2$, we obtain the coupled pair of states with quantum numbers $n=2,3$. Thus, for each $k$, we obtain pairs of states with quantum numbers $n,n+1$. Interestingly, for $k=1$, we \textit{necessarily need} the state with negative energy $-\hbar\omega/2$ to obtain states with  $n=1,2$. Thus, the state with negative energy $-\hbar\omega/2$  can not be neglected. In other words, for small values of ${\cal K}_0>0$, the ground state with energy $\hbar\omega/2$ becomes coupled to the state with energy $-\hbar\omega/2$. \textit{Only such coupling with the state with negative energy} allows to obtain $n=1,2$ when  ${\cal K}_0\to\hbar\omega$. Thus, the prohibited state with negative energy is necessary to obtain the excited states with positive energy in the HO. 

The potential shifts become imaginary and position-dependent for the one-dimensional Hydrogen atom, and ${\cal K}_0$. For the uncoupled case ${\cal K}_0=0$, we obtain two stable (no time dependence) plane-wave-like states at the threshold. For ${\cal K}_0<0$, bounded wave functions can be obtained which are not identical to the usual wave-functions eigenstates but are squeezed closer to the nucleus. For ${\cal K}_0>0$, plane-wave-like solutions are obtained for $\chi^-(q)$,  and $\chi^+(q)$ solutions tends to disappear.

In general, using the classical interpretation for ${\cal P}^{(\pm)}$, namely that
${\cal P}^+>0$ generates space and time evolutions in inverted directions in their axis, and  ${\cal P}^-<0$ generates space and time evolutions in the same directions in their axis. We observe that the vanishing (prohibit) solutions in the quantum dynamics (in case they appear) are always related to ${\cal P}^+$ (\textit{i.e.~}$\chi^+(q)$). This occurs in the HO case and for the continuous states in the one-dimensional Hydrogen atom. For the bounded states of the atom, $\chi^+(q)\ne \chi^-(q)\ne 0$, and are related. Even though $\chi^+(q)$ tend to disappear in some cases, they are needed to obtain $\chi^-(q)$ via coupled Eqs.~(\ref{chi1}) and (\ref{chi2}).

Further developments with this interpretation will consider time-dependent potentials,  higher-dimensional problems and relativistic cases.

\begin{acknowledgments}
We are grateful to Prof JM Rost and Prof RM Angelo for helpful discussions. This study was financed in part by the Coordenação de Aperfeiçoamento de Pessoal de Nível Superior (CAPES, Brazil) – Finance Code  001. MWB thanks Conselho Nacional de Desenvolvimento Científico e Tecnológico (CNPq, Brazil) for financial support (Grant Number: 310294/2022-3,).
\end{acknowledgments}

\appendix

\providecommand{\noopsort}[1]{}\providecommand{\singleletter}[1]{#1}%


\begin{thebibliography}{48}
\expandafter\ifx\csname natexlab\endcsname\relax\def\natexlab#1{#1}\fi
\expandafter\ifx\csname bibnamefont\endcsname\relax
  \def\bibnamefont#1{#1}\fi
\expandafter\ifx\csname bibfnamefont\endcsname\relax
  \def\bibfnamefont#1{#1}\fi
\expandafter\ifx\csname citenamefont\endcsname\relax
  \def\citenamefont#1{#1}\fi
\expandafter\ifx\csname url\endcsname\relax
  \def\url#1{\texttt{#1}}\fi
\expandafter\ifx\csname urlprefix\endcsname\relax\def\urlprefix{URL }\fi
\providecommand{\bibinfo}[2]{#2}
\providecommand{\eprint}[2][]{\url{#2}}

\bibitem[{\citenamefont{Lanczos}(1949)}]{lanczos49}
\bibinfo{editor}{\bibfnamefont{C.}~\bibnamefont{Lanczos}}, ed.,
  \emph{\bibinfo{title}{The Variational Principles of Mechanics}}
  (\bibinfo{address}{Toronto}, \bibinfo{year}{1949}).

\bibitem[{\citenamefont{Synge}(1960)}]{synge60}
\bibinfo{editor}{\bibfnamefont{J.~L.} \bibnamefont{Synge}}, ed.,
  \emph{\bibinfo{title}{Classical Dynamics, in Handbuch der Physik}},
  vol.~\bibinfo{volume}{3} (\bibinfo{address}{Berlin}, \bibinfo{year}{1960}).

\bibitem[{\citenamefont{Haar}(1961)}]{haar61}
\bibinfo{editor}{\bibfnamefont{D.~T.} \bibnamefont{Haar}}, ed.,
  \emph{\bibinfo{title}{Elements of Hamiltonian Mechanics}}
  (\bibinfo{address}{Amsterdam}, \bibinfo{year}{1961}).

\bibitem[{\citenamefont{Hjalmars}(1962)}]{hja62}
\bibinfo{author}{\bibfnamefont{S.}~\bibnamefont{Hjalmars}},
  \bibinfo{journal}{Nuovo Cimento} \textbf{\bibinfo{volume}{XXV}},
  \bibinfo{pages}{354} (\bibinfo{year}{1962}).

\bibitem[{\citenamefont{Grigorescu}(2000)}]{marius00}
\bibinfo{author}{\bibfnamefont{M.}~\bibnamefont{Grigorescu}},
  \bibinfo{journal}{Can.~J.Phys.~} \textbf{\bibinfo{volume}{78}},
  \bibinfo{pages}{959} (\bibinfo{year}{2000}).

\bibitem[{\citenamefont{Dias and Parisio}(2017)}]{diasparisio}
\bibinfo{author}{\bibfnamefont{E.~O.} \bibnamefont{Dias}} \bibnamefont{and}
  \bibinfo{author}{\bibfnamefont{F.}~\bibnamefont{Parisio}},
  \bibinfo{journal}{Phys.\ Rev.\ A} \textbf{\bibinfo{volume}{95}},
  \bibinfo{pages}{032133} (\bibinfo{year}{2017}).

\bibitem[{\citenamefont{Araújo et~al.}(2023)\citenamefont{Araújo, Ximenes,
  and Dias}}]{araujo2023}
\bibinfo{author}{\bibfnamefont{R.~E.} \bibnamefont{Araújo}},
  \bibinfo{author}{\bibfnamefont{R.}~\bibnamefont{Ximenes}}, \bibnamefont{and}
  \bibinfo{author}{\bibfnamefont{E.~O.} \bibnamefont{Dias}},
  \bibinfo{journal}{arXiv} \textbf{\bibinfo{volume}{2306}},
  \bibinfo{pages}{12000} (\bibinfo{year}{2023}).

\bibitem[{\citenamefont{Laskin}(2000{\natexlab{a}})}]{laskin1}
\bibinfo{author}{\bibfnamefont{N.}~\bibnamefont{Laskin}},
  \bibinfo{journal}{Phys. Rev. E} \textbf{\bibinfo{volume}{62}},
  \bibinfo{pages}{3135} (\bibinfo{year}{2000}{\natexlab{a}}).

\bibitem[{\citenamefont{Laskin}(2000{\natexlab{b}})}]{laskin2}
\bibinfo{author}{\bibfnamefont{N.}~\bibnamefont{Laskin}},
  \bibinfo{journal}{Phys. Lett. A} \textbf{\bibinfo{volume}{268}},
  \bibinfo{pages}{298} (\bibinfo{year}{2000}{\natexlab{b}}).

\bibitem[{\citenamefont{Laskin}(2002)}]{laskin3}
\bibinfo{author}{\bibfnamefont{N.}~\bibnamefont{Laskin}},
  \bibinfo{journal}{Phys. Rev. E} \textbf{\bibinfo{volume}{66}},
  \bibinfo{pages}{0506108} (\bibinfo{year}{2002}).

\bibitem[{\citenamefont{Laskin}(2018)}]{laskin}
\bibinfo{editor}{\bibfnamefont{N.}~\bibnamefont{Laskin}}, ed.,
  \emph{\bibinfo{title}{Fractional Quantum Mechanics}} (\bibinfo{address}{World
  Scientific}, \bibinfo{year}{2018}), \bibinfo{edition}{1st} ed.

\bibitem[{\citenamefont{Tarasov}(2013)}]{tarasov2013}
\bibinfo{author}{\bibfnamefont{V.~E.} \bibnamefont{Tarasov}},
  \bibinfo{journal}{Int. Jour. Mod. Phys. B} \textbf{\bibinfo{volume}{27}},
  \bibinfo{pages}{1330005} (\bibinfo{year}{2013}).

\bibitem[{\citenamefont{Tarasov}(2012)}]{tarasov}
\bibinfo{editor}{\bibfnamefont{V.~E.} \bibnamefont{Tarasov}}, ed.,
  \emph{\bibinfo{title}{Fractional Dynamics}} (\bibinfo{publisher}{Springer},
  \bibinfo{address}{London}, \bibinfo{year}{2012}), \bibinfo{edition}{1st} ed.

\bibitem[{\citenamefont{Pauli}(1933)}]{pauli33}
\bibinfo{editor}{\bibfnamefont{W.}~\bibnamefont{Pauli}}, ed.,
  \emph{\bibinfo{title}{Handbuch der Physik, Geiger and Scheel}},
  vol.~\bibinfo{volume}{3} (\bibinfo{year}{1933}).

\bibitem[{\citenamefont{Aharanov and Bohm}(1961)}]{bohm61}
\bibinfo{author}{\bibfnamefont{Y.}~\bibnamefont{Aharanov}} \bibnamefont{and}
  \bibinfo{author}{\bibfnamefont{D.}~\bibnamefont{Bohm}},
  \bibinfo{journal}{Phys.\ Rev.} \textbf{\bibinfo{volume}{122}},
  \bibinfo{pages}{1649} (\bibinfo{year}{1961}).

\bibitem[{\citenamefont{Razavy}(1967)}]{razavy67}
\bibinfo{author}{\bibfnamefont{M.}~\bibnamefont{Razavy}},
  \bibinfo{journal}{Am.~J.~Phys.} \textbf{\bibinfo{volume}{35}},
  \bibinfo{pages}{955} (\bibinfo{year}{1967}).

\bibitem[{\citenamefont{Razavy}(1969)}]{razavy69}
\bibinfo{author}{\bibfnamefont{M.}~\bibnamefont{Razavy}}, \bibinfo{journal}{Il
  N.~Cim.~B~(1965-1970)} \textbf{\bibinfo{volume}{63}}, \bibinfo{pages}{271}
  (\bibinfo{year}{1969}).

\bibitem[{\citenamefont{Allcock}(1969{\natexlab{a}})}]{allcock69-1}
\bibinfo{author}{\bibfnamefont{G.~R.} \bibnamefont{Allcock}},
  \bibinfo{journal}{Ann.~Phys.~} \textbf{\bibinfo{volume}{53}},
  \bibinfo{pages}{253} (\bibinfo{year}{1969}{\natexlab{a}}).

\bibitem[{\citenamefont{Allcock}(1969{\natexlab{b}})}]{allcock69-2}
\bibinfo{author}{\bibfnamefont{G.~R.} \bibnamefont{Allcock}},
  \bibinfo{journal}{Ann.~Phys.~} \textbf{\bibinfo{volume}{53}},
  \bibinfo{pages}{286} (\bibinfo{year}{1969}{\natexlab{b}}).

\bibitem[{\citenamefont{Allcock}(1969{\natexlab{c}})}]{allcock69-3}
\bibinfo{author}{\bibfnamefont{G.~R.} \bibnamefont{Allcock}},
  \bibinfo{journal}{Ann.~Phys.~} \textbf{\bibinfo{volume}{53}},
  \bibinfo{pages}{311} (\bibinfo{year}{1969}{\natexlab{c}}).

\bibitem[{\citenamefont{Kijowski}(1974)}]{kijo74}
\bibinfo{author}{\bibfnamefont{J.}~\bibnamefont{Kijowski}},
  \bibinfo{journal}{Rep.~Math.~Phys.~} \textbf{\bibinfo{volume}{6}},
  \bibinfo{pages}{361} (\bibinfo{year}{1974}).

\bibitem[{\citenamefont{Briggs and Rost}(2000)}]{briggs1}
\bibinfo{author}{\bibfnamefont{J.~S.} \bibnamefont{Briggs}} \bibnamefont{and}
  \bibinfo{author}{\bibfnamefont{J.~M.} \bibnamefont{Rost}},
  \bibinfo{journal}{Eur.\ Phys.\ J.\ D} \textbf{\bibinfo{volume}{10}},
  \bibinfo{pages}{311} (\bibinfo{year}{2000}).

\bibitem[{\citenamefont{Briggs and Rost}(2001)}]{briggs2}
\bibinfo{author}{\bibfnamefont{J.~S.} \bibnamefont{Briggs}} \bibnamefont{and}
  \bibinfo{author}{\bibfnamefont{J.~M.} \bibnamefont{Rost}},
  \bibinfo{journal}{Found.\ of Phys.} \textbf{\bibinfo{volume}{31}},
  \bibinfo{pages}{693} (\bibinfo{year}{2001}).

\bibitem[{\citenamefont{Briggs}(2008)}]{briggs3}
\bibinfo{author}{\bibfnamefont{J.~S.} \bibnamefont{Briggs}},
  \bibinfo{journal}{J.\ Phys.:\ Conf.\ Ser.} \textbf{\bibinfo{volume}{99}},
  \bibinfo{pages}{012002} (\bibinfo{year}{2008}).

\bibitem[{\citenamefont{Brunetti et~al.}(2010)\citenamefont{Brunetti,
  Fredenhagen, and Hoge}}]{hoge10}
\bibinfo{author}{\bibfnamefont{R.}~\bibnamefont{Brunetti}},
  \bibinfo{author}{\bibfnamefont{K.}~\bibnamefont{Fredenhagen}},
  \bibnamefont{and} \bibinfo{author}{\bibfnamefont{M.}~\bibnamefont{Hoge}},
  \bibinfo{journal}{Found.~Phys.~} \textbf{\bibinfo{volume}{40}},
  \bibinfo{pages}{1368} (\bibinfo{year}{2010}).

\bibitem[{\citenamefont{N.~Grot and Tate}(1996)}]{tate96}
\bibinfo{author}{\bibfnamefont{C.~R.} \bibnamefont{N.~Grot}} \bibnamefont{and}
  \bibinfo{author}{\bibfnamefont{R.~S.} \bibnamefont{Tate}},
  \bibinfo{journal}{Phys.~Rev.~A} \textbf{\bibinfo{volume}{54}},
  \bibinfo{pages}{4676} (\bibinfo{year}{1996}).

\bibitem[{\citenamefont{Delgado and Muga}(1997)}]{muga97}
\bibinfo{author}{\bibfnamefont{V.}~\bibnamefont{Delgado}} \bibnamefont{and}
  \bibinfo{author}{\bibfnamefont{J.~G.} \bibnamefont{Muga}},
  \bibinfo{journal}{Phys.~Rev.~A} \textbf{\bibinfo{volume}{56}},
  \bibinfo{pages}{3425} (\bibinfo{year}{1997}).

\bibitem[{\citenamefont{Muga et~al.}(1998)\citenamefont{Muga, Leavens, and
  Palao}}]{muga98}
\bibinfo{author}{\bibfnamefont{J.~G.} \bibnamefont{Muga}},
  \bibinfo{author}{\bibfnamefont{C.~R.} \bibnamefont{Leavens}},
  \bibnamefont{and} \bibinfo{author}{\bibfnamefont{J.~P.} \bibnamefont{Palao}},
  \bibinfo{journal}{Phys.~Rev.~A} \textbf{\bibinfo{volume}{58}},
  \bibinfo{pages}{4336} (\bibinfo{year}{1998}).

\bibitem[{\citenamefont{Egusquiza and Muga}(1999)}]{muga99}
\bibinfo{author}{\bibfnamefont{I.~L.} \bibnamefont{Egusquiza}}
  \bibnamefont{and} \bibinfo{author}{\bibfnamefont{J.~G.} \bibnamefont{Muga}},
  \bibinfo{journal}{Phys.~Rev.~A} \textbf{\bibinfo{volume}{61}},
  \bibinfo{pages}{012104} (\bibinfo{year}{1999}).

\bibitem[{\citenamefont{Muga and Leavens}(2000)}]{muga00}
\bibinfo{author}{\bibfnamefont{J.~G.} \bibnamefont{Muga}} \bibnamefont{and}
  \bibinfo{author}{\bibfnamefont{C.~R.} \bibnamefont{Leavens}},
  \bibinfo{journal}{Phys.~Rep.~} \textbf{\bibinfo{volume}{338}},
  \bibinfo{pages}{353} (\bibinfo{year}{2000}).

\bibitem[{\citenamefont{Anderson}(2012)}]{andersontime}
\bibinfo{editor}{\bibfnamefont{E.}~\bibnamefont{Anderson}}, ed.
  (\bibinfo{publisher}{Nova}, \bibinfo{address}{New York},
  \bibinfo{year}{2012}), \bibinfo{edition}{1st} ed.

\bibitem[{\citenamefont{Galapon et~al.}(2004)\citenamefont{Galapon, Caballar,
  and Bahague}}]{galapon04}
\bibinfo{author}{\bibfnamefont{E.~A.} \bibnamefont{Galapon}},
  \bibinfo{author}{\bibfnamefont{R.~F.} \bibnamefont{Caballar}},
  \bibnamefont{and} \bibinfo{author}{\bibfnamefont{R.~T.}
  \bibnamefont{Bahague}}, \bibinfo{journal}{Phys.~Rev.~Lett.~}
  \textbf{\bibinfo{volume}{93}}, \bibinfo{pages}{180406}
  (\bibinfo{year}{2004}).

\bibitem[{\citenamefont{Galapon
  et~al.}(2005{\natexlab{a}})\citenamefont{Galapon, Caballar, and
  Bahague}}]{galapon05-1}
\bibinfo{author}{\bibfnamefont{E.~A.} \bibnamefont{Galapon}},
  \bibinfo{author}{\bibfnamefont{R.~F.} \bibnamefont{Caballar}},
  \bibnamefont{and} \bibinfo{author}{\bibfnamefont{R.~T.}
  \bibnamefont{Bahague}}, \bibinfo{journal}{Phys.~Rev.~A.~}
  \textbf{\bibinfo{volume}{72}}, \bibinfo{pages}{062107}
  (\bibinfo{year}{2005}{\natexlab{a}}).

\bibitem[{\citenamefont{Galapon
  et~al.}(2005{\natexlab{b}})\citenamefont{Galapon, Delgado, Muga, and
  Egusquiza}}]{galapon05-2}
\bibinfo{author}{\bibfnamefont{E.~A.} \bibnamefont{Galapon}},
  \bibinfo{author}{\bibfnamefont{F.}~\bibnamefont{Delgado}},
  \bibinfo{author}{\bibfnamefont{J.~G.} \bibnamefont{Muga}}, \bibnamefont{and}
  \bibinfo{author}{\bibfnamefont{I.}~\bibnamefont{Egusquiza}},
  \bibinfo{journal}{Phys.~Rev.~A.~} \textbf{\bibinfo{volume}{72}},
  \bibinfo{pages}{042107} (\bibinfo{year}{2005}{\natexlab{b}}).

\bibitem[{\citenamefont{Muga et~al.}(2008)\citenamefont{Muga, Mayato, and
  Gusquiza}}]{secao122}
\bibinfo{editor}{\bibfnamefont{J.~G.} \bibnamefont{Muga}},
  \bibinfo{editor}{\bibfnamefont{R.~S.} \bibnamefont{Mayato}},
  \bibnamefont{and} \bibinfo{editor}{\bibfnamefont{I.~L.}
  \bibnamefont{Gusquiza}}, eds., vol.~\bibinfo{volume}{1}
  (\bibinfo{publisher}{Springer}, \bibinfo{address}{Berlin},
  \bibinfo{year}{2008}), \bibinfo{edition}{2nd} ed.

\bibitem[{\citenamefont{Galapon}(2006)}]{galapon06}
\bibinfo{author}{\bibfnamefont{E.~A.} \bibnamefont{Galapon}},
  \bibinfo{journal}{Int.~J.~.Mod.~Phys.~} \textbf{\bibinfo{volume}{21}},
  \bibinfo{pages}{6351} (\bibinfo{year}{2006}).

\bibitem[{\citenamefont{Galapon and Villaneuva}(2008)}]{galapon08}
\bibinfo{author}{\bibfnamefont{E.~A.} \bibnamefont{Galapon}} \bibnamefont{and}
  \bibinfo{author}{\bibfnamefont{A.}~\bibnamefont{Villaneuva}},
  \bibinfo{journal}{J.~Phys.~A: Math.~Theor.~} \textbf{\bibinfo{volume}{41}},
  \bibinfo{pages}{455302} (\bibinfo{year}{2008}).

\bibitem[{\citenamefont{Galapon}(2009)}]{galapon09}
\bibinfo{author}{\bibfnamefont{E.~A.} \bibnamefont{Galapon}},
  \bibinfo{journal}{Proc.~Roy.~Soc.~A} \textbf{\bibinfo{volume}{71}},
  \bibinfo{pages}{465} (\bibinfo{year}{2009}).

\bibitem[{\citenamefont{Galapon and Magadan}(2018)}]{galapon18}
\bibinfo{author}{\bibfnamefont{E.~A.} \bibnamefont{Galapon}} \bibnamefont{and}
  \bibinfo{author}{\bibfnamefont{J.~J.} \bibnamefont{Magadan}},
  \bibinfo{journal}{Ann.~Phys.~} \textbf{\bibinfo{volume}{397}},
  \bibinfo{pages}{278} (\bibinfo{year}{2018}).

\bibitem[{\citenamefont{Flores and Galapon}(2022)}]{galapon22}
\bibinfo{author}{\bibfnamefont{P.~C.} \bibnamefont{Flores}} \bibnamefont{and}
  \bibinfo{author}{\bibfnamefont{E.~A.} \bibnamefont{Galapon}},
  \bibinfo{journal}{Phys.~Rev.~A} \textbf{\bibinfo{volume}{105}},
  \bibinfo{pages}{062208} (\bibinfo{year}{2022}).

\bibitem[{\citenamefont{de~Lara and Beims}(2023)}]{arlans23-1}
\bibinfo{author}{\bibfnamefont{A.~J.~S.} \bibnamefont{de~Lara}}
  \bibnamefont{and} \bibinfo{author}{\bibfnamefont{M.~W.} \bibnamefont{Beims}},
  \bibinfo{journal}{Phys.Rev.A} \textbf{\bibinfo{volume}{107}},
  \bibinfo{pages}{052220} (\bibinfo{year}{2023}).

\bibitem[{\citenamefont{Ximenes et~al.}(2018)\citenamefont{Ximenes, Parisio,
  and Dias}}]{diasparisioexp}
\bibinfo{author}{\bibfnamefont{R.}~\bibnamefont{Ximenes}},
  \bibinfo{author}{\bibfnamefont{F.}~\bibnamefont{Parisio}}, \bibnamefont{and}
  \bibinfo{author}{\bibfnamefont{E.~O.} \bibnamefont{Dias}},
  \bibinfo{journal}{Phys.\ Rev.\ A} \textbf{\bibinfo{volume}{98}},
  \bibinfo{pages}{032105} (\bibinfo{year}{2018}).

\bibitem[{\citenamefont{P.A.M.Dirac}(1958)}]{dirac}
\bibinfo{author}{\bibnamefont{P.A.M.Dirac}}, \emph{\bibinfo{title}{The
  Principles of Quantum Mechanics}} (\bibinfo{publisher}{Oxford University
  Press}, \bibinfo{year}{1958}), \bibinfo{edition}{4th} ed.

\bibitem[{\citenamefont{Lim et~al.}(2012)\citenamefont{Lim, Eab, Mak, Li, and
  Chen}}]{lim12}
\bibinfo{author}{\bibfnamefont{S.~C.} \bibnamefont{Lim}},
  \bibinfo{author}{\bibfnamefont{C.~H.} \bibnamefont{Eab}},
  \bibinfo{author}{\bibfnamefont{K.~H.} \bibnamefont{Mak}},
  \bibinfo{author}{\bibfnamefont{M.}~\bibnamefont{Li}}, \bibnamefont{and}
  \bibinfo{author}{\bibfnamefont{S.~Y.} \bibnamefont{Chen}},
  \bibinfo{journal}{Math.Prob.Eng.} \textbf{\bibinfo{volume}{2012}},
  \bibinfo{pages}{1} (\bibinfo{year}{2012}).

\bibitem[{\citenamefont{Duan}(2018)}]{duan18}
\bibinfo{author}{\bibfnamefont{J.-S.} \bibnamefont{Duan}},
  \bibinfo{journal}{Advances in Difference Equations}
  \textbf{\bibinfo{volume}{238}}, \bibinfo{pages}{1} (\bibinfo{year}{2018}).

\bibitem[{\citenamefont{J.S.Briggs}(2022)}]{briggsBohm}
\bibinfo{author}{\bibnamefont{J.S.Briggs}}, \bibinfo{journal}{Nat.Sci.}
  p.~\bibinfo{pages}{1} (\bibinfo{year}{2022}).

\bibitem[{\citenamefont{R.Loudon}(1959)}]{loudon59}
\bibinfo{author}{\bibnamefont{R.Loudon}}, \bibinfo{journal}{Am.J.Phys.}
  \textbf{\bibinfo{volume}{27}}, \bibinfo{pages}{649} (\bibinfo{year}{1959}).

\bibitem[{\citenamefont{R.Loudon}(2016)}]{loudon16}
\bibinfo{author}{\bibnamefont{R.Loudon}}, \bibinfo{journal}{Proc.R.Soc.}
  \textbf{\bibinfo{volume}{472}}, \bibinfo{pages}{20150534}
  (\bibinfo{year}{2016}).

\end{thebibliography}

\end{document}